\documentclass[fleqn,usenatbib]{mnras}

\usepackage{units}

\DeclareRobustCommand{\VAN}[3]{#2}
\let\VANthebibliography\thebibliography
\def\thebibliography{\DeclareRobustCommand{\VAN}[3]{##3}\VANthebibliography}
\newcommand{\kpch}{\,h^{-1}\unit{kpc}}
\newcommand{\mpch}{\,h^{-1}\unit{Mpc}}
\newcommand{\hMpc}{\,h\unit{Mpc}^{-1}}
\newcommand{\msunh}{\,h^{-1}\unit{M_\odot}}
\newcommand{\software}[1]{{\small #1}}
\newcommand{\mpgadget}{\software{MP-Gadget}}
\def\astrid{\texttt{Astrid} }
\def\bluetides{\texttt{BlueTides} }

\def\hmpc{h^{-1}{\rm Mpc}}

\def\msun{\, M_{\odot}}

\usepackage{graphicx}	
\usepackage{amsmath}	
\usepackage{amssymb}	

\title[ASTRID Galaxies and Reionization]{The ASTRID Simulation: Galaxy Formation and Reionization}

\author[S.~Bird et al.]{
Simeon Bird,$^{1}$\thanks{E-mail: sbird@ucr.edu}
Yueying Ni,$^{2,3}$
Tiziana Di Matteo,$^{2,3}$
Rupert Croft,$^{2,3}$
Yu Feng,$^{4}$
and Nianyi Chen$^{2}$.
\\
$^{1}$ Department of Physics \& Astronomy, University of California, Riverside, 900 University Ave., Riverside, CA 92521, USA\\
$^{2}$ McWilliams Center for Cosmology, Department of Physics, Carnegie Mellon University, Pittsburgh, PA 15213 \\
$^{3}$ NSF AI Planning Institute for Physics of the Future,
Carnegie Mellon  University, Pittsburgh, PA 15213, USA \\
$^{4}$ Berkeley Center for Cosmological Physics and Department of Physics, University of California, Berkeley, CA 94720, USA
}

\pubyear{2021}

\begin{document}
\label{firstpage}
\pagerange{\pageref{firstpage}--\pageref{lastpage}}
\maketitle

\begin{abstract}
We introduce the \astrid~simulation, a large-scale cosmological hydrodynamic simulation in a $250 \mpch$ box with $2\times 5500^3$ particles. \astrid contains a large number of high redshift galaxies, which can be compared to future survey data, and resolves galaxies in halos more massive than $2\times 10^9 \msun$. \astrid~has been run from $z=99$ to $z=3$. As a particular focus is modelling the high redshift Universe, it contains models for inhomogeneous hydrogen and helium reionization, baryon relative velocities and massive neutrinos, as well as supernova and AGN feedback. The black hole model includes mergers driven by dynamical friction rather than repositioning. We briefly summarise the implemented models, and the technical choices we took when developing the simulation code. We validate the model, showing good agreement with observed UV luminosity functions, galaxy stellar mass functions and specific star formation rates. We show that the redshift at which a given galaxy underwent hydrogen reionization has a large effect on the halo gas fraction. Finally, at $z=6$ halos with $M \sim 2\times 10^9 \msun$ which have been reionized have a star formation rate $1.5$ times greater than those which have not yet been reionized.
\end{abstract}

\begin{keywords}
methods: simulation
--
galaxies:formation
\end{keywords}



\section{Introduction}
With the upcoming launch of the James Webb Space Telescope (JWST) \citep{JWST}, the subsequent Nancy Grace Roman Space  Telescope~\citep[NGRST, formerly WFIRST][]{WFIRST} and the Euclid mission \citep{Euclid2019},
a plethora of new observations of high redshift galaxies will become available \citep[e.g~][]{2020MNRAS.492.5167V}. The massive halos which host bright galaxies become increasingly rare at high redshift as the non-linear scale shrinks, pushing the perturbations that form them ever further from the mean. There have been several recent large scale cosmological simulation projects, such as Magneticum~\citep{Hirschmann2014}, Illustris~\citep{Vogelsberger:2014}, Eagle~\citep{2015MNRAS.446..521S}, Horizon-AGN~\citep{Dubois2015,Volonteri2016Horizon-AGN}, MassiveBlack~\citep{Khandai2015}, BlueTides~\citep{Feng:2016}, Romulus~\citep{Tremmel2017}, Illustris-TNG~\citep{2018MNRAS.475..676S, 2018MNRAS.475..624N}, SIMBA~\citep{Dave2019-simba} or Obelisk \citep{2021A&A...653A.154T}.
Cosmic Dawn \citep{2016MNRAS.463.1462O, 2020MNRAS.496.4087O} and THESAN \citep{2021arXiv211000584K, 2021arXiv211001628G,2021arXiv211002966S} simulated HI reionization using radiative transfer coupled to the hydrodynamics. FLARES in addition used a novel resimulation technique to increase the effective volume simulated \citep{2021MNRAS.500.2127L,2021MNRAS.501.3289V,2021arXiv210800830V}. However, the computational difficulty of simulating increased particle loads mean that existing simulations are either limited to a $100$ Mpc/h box, compromise on the resolution required for following galaxy formation, or make use of resimulation techniques. Making full use of future observations will require a new generation of cosmological simulations with larger boxes \citep{Pillepich:2018}.

Here we present the \astrid~simulation, which models a statistical sample of large halos at the same resolution as other large-scale cosmological galaxy formation simulations for $z > 3$. Concretely, \astrid~has $5500^3$ particles in a $250$ Mpc/h box, achieving a dark matter particle mass resolution of $9.63\times 10^6 M_\odot$. At $z=3$, \astrid contains $7129$ halos with a FOF mass greater than $10^{12} M_\odot$ and $47$ halos with a mass greater than $10^{13} M_\odot$. At $z=6$, it contains $44$ halos with a mass greater than $10^{12} M_\odot$. It thus contains a sample with the statistical power to examine the precursors of Milky-Way like galaxies into the epoch of reionization.

For comparison, the TNG100 simulation had a mass resolution of $7.5 \times 10^6 M_\odot$ in a $75$ Mpc/h box, while TNG300 had a $205$ Mpc/h box with a mass resolution of $5.9\times 10^7 M_\odot$ \citep{2018MNRAS.475..676S, 2018MNRAS.475..624N, 2018MNRAS.480.5113M,2018MNRAS.477.1206N}. The EAGLE simulation had a mass resolution of $9.7 \times 10^6 M_\odot$ in a $70$ Mpc/h box \citep{2015MNRAS.446..521S}. \astrid thus has a mass resolution similar to TNG100 and EAGLE in a box larger than TNG300.

An earlier version of our code was used for the \bluetides simulation. \bluetides studied $z>6$ quasars and galaxies in a $400 \hmpc$ box, and has been validated against a number of observables \citep[e.g.][]{Feng:2016,Wilkins2017,DiMatteo2017,Tenneti2018,Huang2018,Ni2018,Bhowmick2018,Ni2020,Marshall2020,Marshall2021}.
\astrid~has a moderately lower total particle load than BlueTides ($2\times 5500^3$ vs $2 \times 7040^3$). However, it improves on the BlueTides mass resolution of $1.7 \times 10^7 M_\odot$ by almost a factor of two. Most importantly, for computational reasons the lowest redshift reached by BlueTides was $z=6.5$, but \astrid~reached $z=3$.

The primary scientific focuses of \astrid~are high redshift galaxy formation and black hole mergers. To improve modelling of these areas, we have included several new physical processes into our simulation code. Importantly for galaxy formation, we include models for inhomogeneous hydrogen reionization (Section~\ref{sec:hydrogen}), helium reionization (Section~\ref{sec:helium}), metal return from massive stars (Section~\ref{sec:metals}), and the initial velocity offset between baryons and dark matter (Section~\ref{sec:ics}). To better describe black hole mergers, we have implemented a dynamic friction model (Section~\ref{sec:blackhole}). \astrid~was run using $3\times 10^6$ node-hours on the Frontera supercomputer (where each node contains $56$ Intel Xeon Platinum 8280 cores) using the \mpgadget~code \citep{MPGadget2018}. \mpgadget~implements a hybrid parallelization model with extensive shared and distributed memory parallelism.

In addition to describing the techniques used in performing the simulation, we present some initial results for the population of high redshift galaxies, compared to available observations in Section~\ref{sec:results}. Results for the black hole population may be found in our companion paper \citep{Ni:2021inprep}.

\begin{table*}
  \begin{tabular}{ccl}
  \hline
    Name & Value & Notes \\
    \hline
    $L_\mathrm{Box} $ &     $250\mpch$ & Simulation box size. \\
    $N_\mathrm{DM}$ &     $5500^3$  &   Number of dark matter particles. \\
    $N_\mathrm{Baryon}$ &     $5500^3$  &   Initial number of gas particles. \\
    $M_\mathrm{DM}$  &     $6.74\times10^6 \msunh$  &   Mass of a dark matter particle. \\
    $M_\mathrm{GAS}$  &    $1.27\times10^6 \msunh$ &     Mass of a gas particle in the initial conditions. \\
    $z_\mathrm{init}$ & $99$ & Initial redshift \\
    $z_\mathrm{final}$ & $3$ & Final redshift \\
    $\Omega_\Lambda$  & $0.6911$  & Dark energy density \\
    $\Omega_\mathrm{0}$ & $0.3089$ &   Total matter density\\
    $\Omega_\mathrm{Baryon}$ &    $0.0486$  & Baryon density\\
    $M_\nu$ & $0.06$ eV & Total neutrino mass \\
    $h$  &  $0.6774$ & Hubble parameter in units of $100\,\unit{km/s/\mathrm{Mpc}}$ \\
    $A_\mathrm{s}$  &      $2.142\times 10^{-9}$ & Primordial amplitude of scalar perturbations.\\
    $\sigma_8(z=0)$ &  $0.816$ & Present day fluctuation amplitude on $8$ Mpc scales. \\
    $n_s$  &     $0.9667$ & Scalar spectral index. \\
    $\varepsilon_\mathrm{grav}$ &    $1.5\kpch$ &     Gravitational softening length. \\
    $N_\mathrm{Generation}$ &     $4$ & Average number of star particles produced per gas particle. \\
    $\mathrm{egy}_w$ &      $1.0$  &  Supernova feedback energy in units of $10^{51}\,\unit{erg/s}$\\
    $\kappa_w$  &     $3.7$ &    Wind speed in units of local dark matter velocity dispersion. \\
    $\eta_\mathrm{BH}$ &    $0.05$ &    Black hole feedback efficiency. \\
    $M^{(0)}_\mathrm{BH,min}$ &    $3\times 10^{4}\msunh$ &   Minimum black hole seed mass.\\
    $M^{(0)}_\mathrm{BH,max}$ &    $3\times 10^{5}\msunh$ &  Maximum black hole seed mass.\\
    $M^{(0)}_\mathrm{HALO}$ &    $5\times10^{9}\msunh$ &    Minimum halo mass to host a black hole seed.\\
    $M^{(0)}_\mathrm{STAR}$ &    $2\times10^{6}\msunh$ &    Minimum stellar mass required for black hole seed.\\
    $z_{re}$ & $7.5$ & Midpoint of hydrogen reionization. \\
    $z_{HeII}$ & $4.4 - 2.8$ & Redshift range of helium reionization. \\
    $\alpha_q$ & $2.0$ & Helium reionization heating parameter. \\
    Seed & $9281110$ & Initial GenIC structure seed. \\
    \hline
  \end{tabular}
  \caption{Summary of the main parameters of the \astrid~simulation. Star formation, SPH and some black hole parameters are set to their respective defaults.}
  \label{tab:simparam}
\end{table*}

\section{Methods}
\label{sec:methods}

\mpgadget~\citep{MPGadget2018} is a massively scalable version of the cosmological structure formation code Gadget-3 \citep{Springel:2005}. The main new feature of \mpgadget~over Gadget-3 is shared memory parallelism using OpenMP threads, in addition to distributed memory parallelism using MPI. Gadget divides space into contiguous regions called processor domains by subdividing a Peano-Hilbert curve. Particles found within the contiguous region(s) assigned to a specific MPI rank are stored in local memory, while other particles are found in the memory assigned to their respective MPI ranks. \mpgadget~maintains a tree structure which covers all processor domains and which is walked to determine when communication is necessary. Our implementation of shared memory parallelism does not change this picture, but it makes the characteristic size of a domain larger, by associating all processors on a socket with a single MPI rank.

Thus, while the basic algorithms are mostly unchanged from Gadget-3, the evolution from a pure MPI code has resulted in substantial changes to their implementation. In this Section we detail the main changes from \cite{Feng:2016} to our physics algorithms and the continuing improvements to scalability. 
We continue to use the \software{BigFile}\footnote{\url{http://github.com/rainwoodman/bigfile}} snapshot format, which transparently supports file-level striping and allows our IO to fully saturate Frontera's IO servers.

\subsection{Gravity}
\label{sec:gravity}
The core of \mpgadget~remains a TreePM gravity solver \citep{Xu:1995, Bagla:2002, Springel:2005}, as in Gadget-3. Long-range forces are computed on a particle mesh in Fourier space, while short-range forces are computed using a hierarchical force tree. The short-range gravitational solver in particular has seen substantial scalability improvements which enabled an increase in the default force accuracy. We have verified that with our settings \mpgadget~is able to reproduce the matter power spectrum output by PKDGRAV3 \cite{2016JCAP...04..047S} with an error less than $1\%$.

Every particle-mesh step we shift the positions of the particles by a constant random offset chosen uniformly between $0$ and $8$ particle-mesh grid cells, transparently undoing this offset before writing snapshots. This randomization of the particle positions relative to the box has no dynamic effect in a periodic box. However, it avoids temporal correlations in the tree opening criterion, which can eventually lead to pathological errors in tree node boundaries at high redshift, as explained in \cite{2020arXiv201003567S}.

\subsubsection{Long-range gravity: particle mesh}

Long-range gravitational forces are computed using a particle mesh approach. A density grid is computed using a cloud-in-cell mass deposition. The gravitational field is computed using the appropriate force kernels and real space forces are computed using Fourier-space finite differencing \citep[as in the HACC code:][]{2014arXiv1410.2805H}.

Gadget-3's default particle mesh gravity solver
(used to compute long-range gravitational forces) is based on FFTW2. The distributed memory implementation of FFTW divides data into chunks along the first dimension of the transform and thus a 3D $N^3$ Fourier transform can only be distributed onto $N$ processors, a scalability limitation. \mpgadget~ instead uses a gravity solver based on \software{PFFT} \citep{doi:10.1137/120885887}, which divides data a 3D transform along $2$ dimensions, allowing the transform to be distributed among $N^2$ processors.

In concert these changes removed the particle mesh step as a scalability bottleneck. The size of the particle grid used in \astrid~is $2^3\times N_\mathrm{PART} = 11000^3$, limited by available memory rather than the scalability of the Fourier transforms.

\subsubsection{Short-range gravity: multipole tree}

Short-range forces are computed using a hierarchical multipole expansion of the gravitational field, leading to a uniformly high force resolution throughout the computational volume. As in Gadget-3, this short-range force is computed by walking a distributed tree structure, each layer of which contains moments describing the total mass within the node and the node center of mass \citep{1986Natur.324..446B}.\footnote{We do not implement the Fast Multipole Method (FMM) for short-range forces. \cite{2020arXiv201003567S} showed a relatively low performance improvement for FMM-PM over TreePM.} The short-range force algorithm is faster than a direct summation because the forces from a large number of distant particles can often be replaced by the force from a single multipole with minimal loss of accuracy.

Gravitational forces are computed by walking the moment tree for each active particle. The local tree is walked for each particle first, with remote nodes being added to a communication list and walked in a second step. Shared memory parallelism is implemented by splitting particles into batches, with each batch processed in parallel by a different thread. The chunk size is initially $100$ particles, and is reduced in factors of $2$ when the number of remaining particles is lower than the current chunk size, as in the `guided' OpenMP scheduling strategy.

\subsubsection{Building the Gravity Tree}

In \astrid, the tree structure and associated moments are rebuilt every timestep. This differs from the version of the code used in \bluetides, where the tree persists and particle motion is accounted for by moving tree nodes according to the average velocity of particles within each node. However, we found that the \bluetides implementation limited force accuracy, as the velocity of individual particles may deviate substantially from the average in the node. Scalability was also adversely impacted, as the code was unaware if a particle crossed into a part of the particle tree on a separate processor, causing excessive communication between the stranded particle and the neighbours \citep{Bird:2018}.

Since the tree is rebuilt every timestep, it represents a calculation overhead which limits the overall simulation time for deep timestep hierarchies. The tree build must thus be highly optimised and scale well with the number of threads. Building the tree has three separate substeps: in the first a shallow tree is built with leaf nodes, called top nodes, corresponding to the processor domains. In the second, particles are added to this shallow tree incrementally, with layers of extra tree nodes created once a leaf node contains the maximum number of particles ($8$). The third step is to compute the tree moments: the total mass and center of mass of all particles in the subtree. The first step has a trivial computational cost, due to the low number of nodes involved.

The second step, tree node creation, is the most expensive part of the tree build and the most involved to parallelise. Our initial parallel implementation, used in \cite{Bird:2018}, guarded each tree node with a spinlock. New particles were added to the tree with this lock taken. Careful use of atomic instructions allowed us to walk the tree and find the right location to add a particle, taking the lock only when the correct node was identified. However, for \astrid, we found that the cost of the atomic instructions limited scalability to $\sim 10$ threads, less than half the hardware optimum of $28$.

We have thus implemented a new, lockless, tree build algorithm. Each processor starts with a copy of the top nodes for the local MPI rank. Particles are processed in a parallel loop, with each particle added to the tree local to each processor. Once all particles have been processed, the partial trees are merged. Note that it is not possible a priori to ensure that these trees are disjoint. However, the particles are usually close to being in Peano-Hilbert order and so most of the particles associated with a particular top node will be spread over $1-2$ processors. Merging is done by simultaneously walking two particle trees, from a top node. When one of the walks encounters a leaf node, the particles in this leaf node are merged into the larger subtree and the parent nodes are updated to point to the new larger subtree. We create a copy of the top nodes for each of the $N_t$ OpenMP threads. To ensure our merging is lockless, we only allow one thread to merge together these $N_t$ top node copies. We also over-decompose the top-level domain to one top node per OpenMP thread (rather than one per MPI rank), so that all threads are busy during the merge step.

Finally, moments are computed recursively, using almost the same algorithm as in Gadget-3. Computation starts at each top node, stopping at leaf nodes. OpenMP $4$ tasks are spawned at each recursion level, to a maximum of $512$ tasks. As a further performance optimisation we compute the moments for leaf nodes during the initial addition of the particle, thus avoiding the need to read particle data a second time during moment computation.

The tree rebuild still represents a moderate overhead, averaging about $25\%$ of the total runtime for \astrid, with an average number of $11$ active timesteps. A future direction towards further reducing this overhead would be to implement the Hamiltonian gravitational timestep splitting from \cite{2020arXiv201003567S}. This would avoid the need to rebuild the gravitational moments, although SPH neighbour finding would still require a tree for the gas particles.

\subsubsection{Long/Short-range Force Split}

Since short and long range forces are computed separately in a TreePM code, it is necessary to match them together. To reduce force anisotropy due to the PM grid, the smoothing scale is chosen to be moderately larger than the size of a single grid cell. Thus both the long-range and short-range gravitational potentials are smoothed. The long-range force is smoothed with an exponential:
\begin{equation}
    \phi_\mathrm{long} = \phi \exp\left( -k^2 r_s^2\right)\,,
\end{equation}
where $r_s$ is the force split scale and we define $A_\mathrm{smth} = r_s / (L / N_\mathrm{grid})$ to be the force split in units of the size of a single PM grid cell. $N_\mathrm{grid}$ is the linear number of PM cells per side and $L$ is the box size. The smoothed short-range force is evaluated to a distance $R_{\mathrm{cut}} r_s$. Both $A_\mathrm{smth}$ and $R_{\mathrm{cut}}$ are parameters of the method; higher parameters are more accurate but require more force evaluations. We have increased $A_\mathrm{smth}$ from $1.25$ to $1.5$ and $R_{\mathrm{cut}}$ from $4.5$ to $6$. We have replaced the error function approximation to the short-range smoothing force with a numerical smoothing function computed by evaluating the average forces from a smoothed PM grid. With these settings, force accuracy as tested against an exact N-body solver is below $1\%$ \citep{2020arXiv201003567S}.

\subsubsection{Massive Neutrinos}
\label{sec:neutrinos}

MP-Gadget includes a fast and accurate model for the effect of massive neutrinos on cosmological structure, following \cite{Yacine:2013}. This model includes the dynamic effect of neutrinos in each timestep.
No explicit neutrino particles are included, but the neutrino perturbation is computed on the PM grid, using a linear response formula, from the non-linear growth of the dark matter. Short-range neutrino forces are neglected, a good approximation as neutrinos are free-streaming on scales smaller than a PM cell.
We assume a single massive neutrino species with a mass of $0.06$~eV, the minimum from neutrino oscillation experiments \citep{deSalas:2018}. Neutrinos of this mass suppress the matter power spectrum by about $7-8\%$ at $k = 1 \hMpc$ and $z=9 - 2$ \citep{Bird:2018}.

\subsection{Initial Conditions}
\label{sec:ics}

Initial conditions for \astrid~are generated at $z=99$ using a power spectrum output by CLASS \citep{CLASS}, Cosmological parameters follow \cite{Planck}, and are given in Table~\ref{tab:simparam}. Uniquely for cosmological hydrodynamic simulations of this size, the initial conditions are generated using separate transfer functions for the cold dark matter and baryons, following \cite{Bird:2020}. These separate transfer functions, evaluated on small scales, cause a stochastic scatter in the relative velocities of CDM and baryons \citep{2010PhRvD..82h3520T}. On average, this effect suppresses star formation in small halos at $z> 10$ \citep[e.g~][]{Naoz:2007}. Note that these velocity fluctuations are sourced from scales of $> 100$ Mpc, and so \astrid's large box is necessary to correctly model them.

The initial distribution of cold dark matter particles is a grid, and the initial distribution of baryons is a Lagrangian glass, as recommended by \cite{Bird:2020} to reduce particle transients. First-order Lagrangian perturbation theory \citep{Zeldovich:1970, Crocce:2006} is used to initialise particle positions and velocities. Second-order terms are not included as they are difficult to model when the initial transfer functions for baryons and cold dark matter are different. Radiation density is included in the cosmological background evolution, a $10\%$ contribution to the Hubble expansion at $z=100$ that affects the total integrated growth rate at the $10\%$ level by $z=10$.

\subsection{Hydrodynamics}

We adopt the pressure-entropy formulation of smoothed particle hydrodynamics (pSPH) to solve the Euler equations \citep{2013MNRAS.428.2840H,2010MNRAS.405.1513R} as implemented by \cite{2014MNRAS.440.1865F}. The density estimator uses a quintic density kernel to reduce noise in SPH density and gradient estimation \citep{2012JCoPh.231..759P}.

Following \cite{2020arXiv201003567S} we incorporate an extra timestep criterion to limit the rate of change of the SPH smoothing length:
\begin{equation}
    \Delta t_i = C_\mathrm{CFL} h_i\left(\frac{\mathrm{d} h_i}{\mathrm{d}t}\right)^{-1}\,.
\end{equation}
The Courant factor $C_\mathrm{CFL} = 0.15$. This criterion affects $\sim 1/10^{4}$ gas particles, and has minimal dynamic impact, but it prevents the smoothing length estimate becoming too inaccurate for inactive particles.

The SPH artificial viscosity is computed as a function of the relative velocity between a particle and its neighbours. Thus when a neighbour particle is not active, the effect of a partial kick to the current time of the active particle must be predicted. Similarly, the effect on the internal energy of the inactive particle of a partial cooling step should be applied when computing artificial viscosities. We found that computing these predicted quantities for every particle during every timestep represented a large source of overhead when the number of active particles was small. Accordingly, we compute predicted velocities and internal energies lazily. Predicted quantities are initially zero on each timestep. The first time a neighbour search encounters a particular particle, it initialises the predicted velocity using an atomic write.

\subsection{Hydrogen Reionization}
\label{sec:hydrogen}

\begin{figure*}
\centering
  \includegraphics[width=0.45\textwidth]{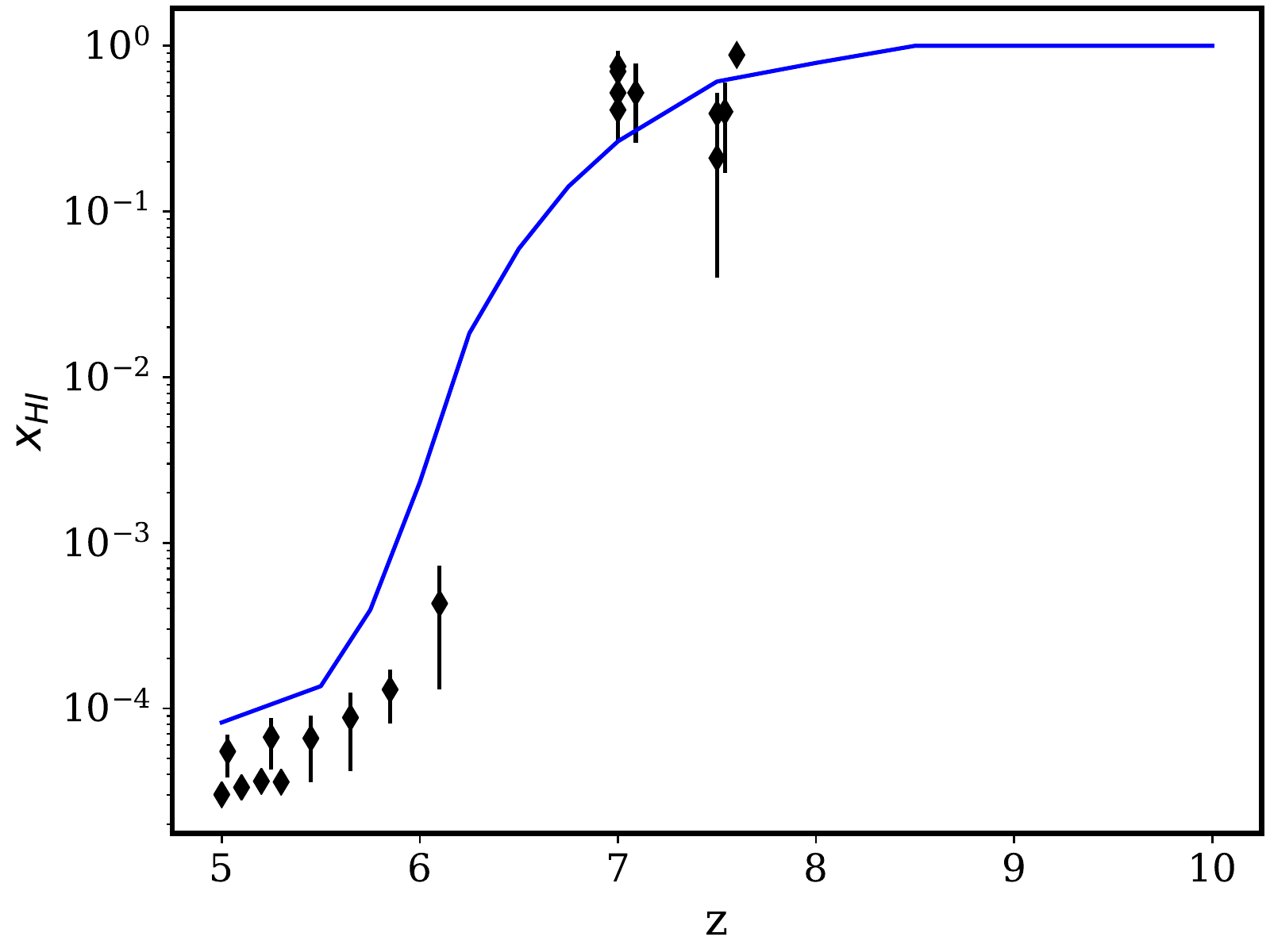}
    \includegraphics[width=0.45\textwidth]{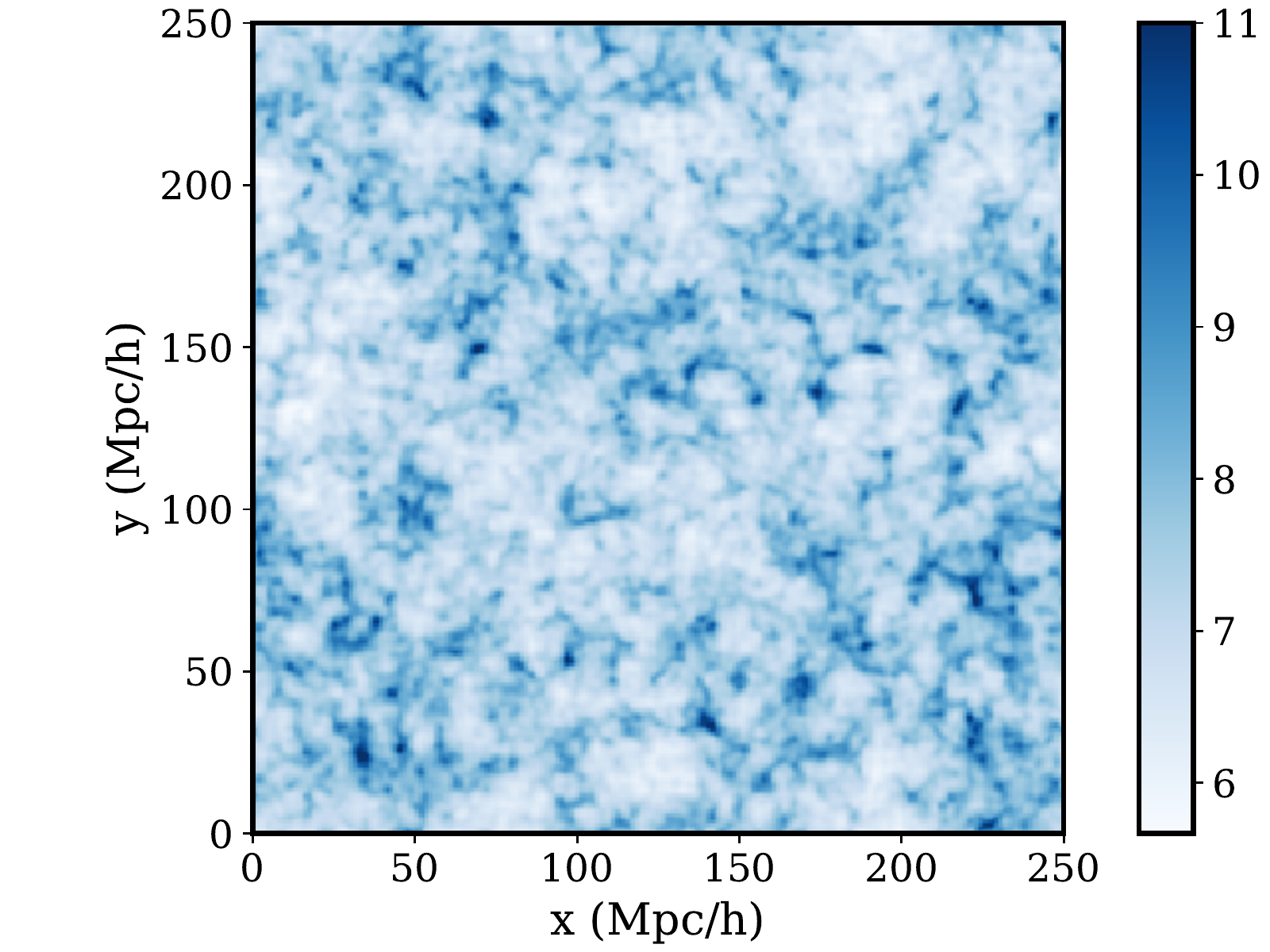}
  \caption{HI reionization history as a function of redshift. Shown is the mean volume-weighted neutral fraction estimated from $10000$ randomly chosen gas particles. (Right) Slice showing reionization redshift (ie, redshift at which each cell is first exposed to the UVB) in cells of depth $0-1$ Mpc/h.  The reionization midpoint is at $z=7.5$.  Observational data is included from \protect\cite{Fan_2006, Ota_2008, Ono_2011, Davies_2018, 2019MNRAS.485.3947M, Greig_2019, Hoag_2019, 2020ApJ...896...23W, 2020ApJ...897L..14Y, Bosman:2021}.}
  \label{fig:reion_hist}
\end{figure*}

We model patchy reionization with a spatially varying ultra-violet background using a semi-analytic method based on hydrodynamic simulations performed with radiative transfer \citep[for more details see][]{2013ApJ...776...81B}. The radiative transfer simulation is used to correlate reionization redshift with overdensity:  larger overdensities are assumed to contain more sources and hence reionize earlier. We generate a reionization redshift field in advance of the simulation using an approximate overdensity field computed from the initial conditions using FastPM \citep{2016MNRAS.463.2273F}. The reionization redshift is stored on a grid with a resolution of $1$ Mpc/h. During the simulation, each particle's position is checked against this grid to determine the reionization redshift.

If the current time is earlier than this redshift, the photon background is set to zero and the particle remains neutral. In regions that have been reionized, we assume the UV background estimated by \cite{2020MNRAS.493.1614F} (note this is zero for $z>10$). We use the optically thin version, as the attenuating effects of reionization are included by our explicit reionization model.

Our fiducial reionization model has a median reionization redshift of $z \sim 7.5$, as suggested by the optical depth measurement of \cite{Planck}. Figure \ref{fig:reion_hist} shows the reionization history as a function of redshift. Although small parts of the box reionize early, the photon background is assumed to be zero at $z>10$ so this has no dynamic effect. Note that $10\%$ of the volume is neutral at $z=6.5$ and a small tail of gas remains neutral until $z < 6$. Shown is the mean volume-weighted neutral fraction estimated from $10000$ randomly chosen gas particles\footnote{We have checked our results are insensitive to this number}. We are in good agreement with the observations at $z > 7$, but have a high neutral fraction for $z < 6$. By this redshift, the neutral fraction is dominated by self-shielded gas around galaxies as reionization is largely complete. This discrepancy thus indicates that the strength of the UVB may be somewhat underestimated at these redshifts. The right panel of figure~\ref{fig:reion_hist} shows the reionization redshift in a 2D slice of the box, the morphology of reionization in our model.

\subsection{Helium Reionization}
\label{sec:helium}

We additionally include a model for spatially inhomogeneous helium reionization following \cite{2020MNRAS.496.4372U}. Helium reionization begins at $z=4.4$ and finishes at $z=2.8$. The progress of helium reionization and its topology is sensitive to the uncertain properties of the quasar population, beaming and quasar lifetime, and depends on redshift \citep{2016ApJ...828...90L, 2017ApJ...841...87L,2018ApJ...868..106L}. Since our simulation does not include radiative transfer, we model the topology of helium reionization with a simple ansatz, motivated by radiative transfer simulations. We assume a linearly increasing reionized volume fraction. Reionized bubbles are placed around randomly chosen halos with mass $M_{\rm halo}\geq10^{12}$M$_{\odot}$, chosen to have a reasonable chance of hosting a quasar. Bubbles are $30$ Mpc/h across, matching the mean bubble size found in radiative transfer simulations \citep{2009ApJ...694..842M}. Inside this bubble a quasar radiation field is assumed with a spectral index of $\alpha_q = -2$. The bubble size is large compared to the galaxy scale, so these choices should not strongly affect galaxy formation.

The model creates a split between photons with a mean free path larger than this bubble size, whose heating is handled homogeneously, and photons with a mean free path shorter than this bubble size, which instantaneously ionize the intergalactic gas. Particles are marked as ionized, and abruptly heated. Once marked they see the same homogeneous UVB as would be visible with the helium reionization model off.

Our model includes the patchiness of helium reionization and the resulting fluctuations in temperature, ionization state, and pressure smoothing. It captures the true scatter in the intergalactic medium temperature-density relation that should result from \ion{He}{II} reionization \citep{2020MNRAS.496.4372U}.

\subsection{Star Formation}

The star formation model is unchanged from \cite{Feng:2016}, itself an implementation of \cite{2003MNRAS.339..289S}. Gas is allowed to cool both radiatively \citep{1996ApJS..105...19K} and via metal line cooling. We include self-shielding of dense gas via the fitting function of \cite{Rahmati:2013}. We approximate the metal cooling rate by scaling a solar metallicity template according to the metallicity of gas particles, following \cite{Vogelsberger:2014}.
We include a correction for the formation of molecular hydrogen, and its effect on star formation at low metallicities, according to the prescription by \cite{2011ApJ...729...36K}. Stars are formed with $1/4$ of the mass of a gas particle. Since we have also implemented mass return, the mass of a gas particle after forming $4$ stars may exceed zero. To prevent runaway enrichment as gas cycles between a star and gas particle, the $5$th star spawned from a gas particle contains the entire mass of the particle.

\subsection{Supernovae Winds and Stellar Feedback}

A stellar wind feedback model \citep{2010MNRAS.406..208O} is included, which assumes wind speeds proportional to the local one dimensional dark matter velocity dispersion $\sigma_\mathrm{DM}$:
\begin{equation}
v_w = \kappa_w \sigma_\mathrm{DM} \,,
\end{equation}
where $v_w$ is the wind speed. $\kappa_w$ is a dimensionless parameter, which we take to be $3.7$ following \cite{Vogelsberger:2013}.
Winds are sourced by newly formed star particles, which randomly pick gas particles from within their SPH smoothing length to become wind particles. The total mass loading is $(v_w/ 350 \mathrm{km/s})^{-2}$ where $350$ km/s is in physical units. Once a particle is in the wind, it is hydrodynamically decoupled for the minimum of $60$ Myr or $20 \mathrm{kpc} / v_w$. Particles are also recoupled when their density drops by a factor of $10$, a condition which dominates most of the wind particles. Particles in the wind do not experience or produce pressure forces, nor may they be accreted onto a black hole. However, they do receive mass return, cool, and contribute to density estimates. There have been a number of improvements to the wind model implementation to ensure that the model is deterministic when executed in a threaded environment. In previous versions, if two newly created star particles were able to kick the same gas particle into the wind, which one of them performed the kick was chosen at random. Now the star particle closest to the kicked gas particle is chosen. We implemented a maximum wind decoupling time, following \cite{2018MNRAS.475..624N}. Decoupled particles are included when estimating the gas density, although not for the black holes, to prevent the gas density varying dramatically over a single timestep. The overall effect of these changes is that stellar feedback is more effective in \astrid than in \bluetides.

\subsection{Metal Return}
\label{sec:metals}

We include a model for the return of mass and metal to the inter-stellar medium from massive stars. The general approach follows \cite{Vogelsberger:2013, Pillepich:2018}, although the implementation is independent and we have created our own metal yield tables. Each star particle is treated as a single stellar population, with the proportion of massive stars set by a Chabrier initial mass function (IMF) \citep{Chabrier:2003}. Massive stars have a finite lifetime, set using the lifetime tables of \cite{Portinari:1998}, and at the end of this lifetime they return mass and metal to neighbouring gas particles proportional to their fraction of the underlying stellar population. Nine metal species are followed: H, He, C, N, O, Ne, Mg, Si, Fe. The total metallicity is tracked separately.

Metal and mass yields from AGB stars, SnII and Sn1A are pre-compiled and included as a function of mass and stellar metallicity. Stars with masses $1 M_\odot - 8 M_\odot$ are assumed to return via an Asymptotic Giant Branch (AGB) stage, with yields taken from \cite{Karakas:2010}, supplemented with \cite{Doherty:2014a, Doherty:2014b} for $M > 6.5 M_\odot$. Masses $8 M_\odot - 40 M_\odot$  return via SnII, with yields from \cite{Kobayashi:2006}. For $8 M_\odot < M < 13 M_\odot$, the tables do not contain data and so we use the metal yields for $13 M_\odot$ stars scaled by $M / (13 M_\odot)$. Sn1a yields follow the W7 model of \cite{Nomoto:1997}. The main difference between our model and that of \cite{Pillepich:2018} lies in the treatment of stars with masses $8 - 13 M_\odot$, where \cite{Pillepich:2018} use a combination of the tables from \cite{Portinari:1998} and \cite{Karakas:2010}.

Mass and metal return is processed only when a star particle yields more than $10^{-3}$ of its initial mass in a single timestep. An SPH-like density kernel is computed for the star particle. Mass and metals are added to all gas particles within this kernel, weighted by gas particle volume and star particle SPH kernel. A technical difference between our model and that of \cite{Pillepich:2018} is that our SPH-based hydrodynamic solver lacks a mechanism for particle splitting. This reduces (numerical) diffusion of metals, but leads to the possibility of large mass variation in gas particles inside the most massive star forming halos. To avoid this we cap the gas particle mass after stellar return: any mass returned to a particle in excess of $4$ times the initial gas mass is assumed to be retained within the star. As most overweight gas particles quickly form stars, the fraction of gas particles affected at any one time is extremely low, $\sim 1/(5\times 10^7)$ in \astrid.

\subsection{Black Holes}
\label{sec:blackhole}

We include a treatment of super-massive black holes (SMBHs). The SMBH model is discussed in more detail in \cite{Ni:2021inprep} and summarised here. Our accretion and feedback models are similar to those in the \bluetides simulation \cite{Feng:2016}, and are based on earlier work by \cite{SDH2005,DSH2005}. The radiative efficiency of feedback is $0.05$.

We have modified the SMBH seeding scheme, drawing the BH seed mass from a power-law distribution instead of using a universal seed mass. SMBH particles are seeded by converting the densest gas particle. To be eligible for seeding, a halo must have total mass greater than $5\times10^{9}\,h^{-1}\unit{M_\odot}$ and stellar mass greater than $2 \times 10^6 h^{-1} M_\odot$. These values are chosen so that BHs are seeded only in halos with sufficient stellar mass for the dynamic friction to be effective.
BH seed masses are drawn probabilistically from a power-law distribution with index $-1$, minimum mass $3 \times 10^4 h^{-1} M_\odot$ and maximum mass $3 \times 10^5 h^{-1} M_\odot$, chosen to match extrapolations of the black hole mass function. Considering the complex astrophysical processes involved in SMBH seeding, a power-law distribution is likely more realistic, and allows us to avoid any effect from harmonics of a single seed mass. \cite{Ni:2021inprep} discuss the effect of the seed mass in \astrid in more detail.

Instead of constantly repositioning the black hole towards the potential minimum, as in earlier simulations, we implement a model for dynamical friction \citep{Tremmel2015, Tremmel2017, Chen2021}. Dynamical friction is an artificial force modelling unresolved small-scale interactions between the SMBH and nearby stars. These interactions transfer momentum from the SMBH to individual stars in the surrounding star clusters,  gradually reducing the momentum of the SMBH particle relative to the surrounding collisionless objects in the bulge \citep[e.g.][]{Governato1994,Kazantzidis2005}. Dynamical friction also stabilizes the BH motion at the center of the galaxy. We estimate dynamical friction using Eq. 8.3 of \cite{Binney2008}:
\begin{equation}
\label{eq:df_full}
    \mathbf{F}_{\rm DF} = -16\pi^2 G^2 M_{\rm BH}^2 m_{a} \;\text{log}(\Lambda) \frac{\mathbf{v}_{\rm BH}}{v_{\rm BH}^3} \int_0^{v_{\rm BH}} dv_a v_a^2 f(v_a)\,.
\end{equation}
$M_{\rm BH}$ is the BH mass, $\textbf{v}_{\rm BH}$ is the BH velocity relative to its surrounding medium, $m_a$ and $v_a$ are the masses and velocities of the particles surrounding the BH. The Coulomb logarithm $\Lambda$ is calculated with
\begin{equation}
    \Lambda = \frac{b_{\rm max}}{(GM_{\rm BH})/v_{\rm BH}^2}\,,
\end{equation}
where $b_{\rm max} = 20\text{ kpc}$.
$f(v_a)$ in Eq.~\ref{eq:df_full} is the velocity distribution of the surrounding collisionless particles including both stars and dark matter, which we approximate by a Maxwellian distribution for each particle.
Eq.~\ref{eq:df_full} reduces to
\begin{equation}
    \label{eq:H14}
    \mathbf{F}_{\rm DF} = -4\pi \rho_{\rm sph} \left(\frac{GM_{\rm BH}}{v_{\rm BH}}\right)^2  \;\text{log}(\Lambda) \mathcal{F}\left(\frac{v_{\rm BH}}{\sigma_v}\right) \frac{\bf{v}_{\rm BH}}{v_{\rm BH}}.
\end{equation}
Here $\rho_{\rm sph}$ is the density of dark matter and star particles within the SPH kernel.
The function $\mathcal{F}$, defined as
\begin{equation}
    \label{eq:fx}
    \mathcal{F}(x) =  \text{erf}(x)-\frac{2x}{\sqrt{\pi}} e^{-x^2},
\end{equation}
is the result of analytically integrating the Maxwellian distribution. $\sigma_v$ is the velocity dispersion of the surrounding particles.

\subsection{Friends of Friends Groups}
\label{sec:fof}

We define dark matter halos primarily using the friends of friends (FOF) algorithm \citep{Davis1985}. This algorithm progressively associates each dark matter particle with its neighbours. Particles are associated when they are within the linking length ($0.2 \times$ mean dark matter particle separation: $9$ kpc/h for \astrid). We also associate gas, star and black hole particles with their nearest dark matter particle (subject to a maximum of $4\times$ the linking length) and thus the host halo.

We post-processed some snapshots with \texttt{SUBFIND} \citep{2001MNRAS.328..726S} in order to identify substructure. At the relatively high redshifts of the simulation, we found that the largest \texttt{SUBFIND} mass is similar to the total FOF mass. We found \texttt{SUBFIND} difficult to scale to the full simulation volume and so instead ran it on separated chunks of the FOF particle tables , with chunk boundaries aligned to the edges of FOF halos, so that \texttt{SUBFIND} is always run on complete halos, and $\sim 512^3$ particles for each chunk.


\subsection{Scalability and Performance}

\subsubsection{Load Balancing}
\label{sec:loadbalance}

We found that for test hydrodynamical simulations, work-load balancing based on the number of short-range gravitational interactions became less effective than simply balancing for equal particle number. Although the nominal work imbalance was indeed reduced, the total compute time to complete the simulation was $10\%$ longer. In \astrid we decomposed for equal particle load per MPI rank.

We ascribe this seemingly counter-intuitive result to three factors.
First, with OpenMP threads the size of the particle domain on each MPI rank contains more particles and is thus spatially larger. The average spatial region covered by an MPI rank is $(15 \mathrm{Mpc/h})^3$. On these scales structures are almost linear and overdensities are small, reducing the potential benefit of work-load balancing. Second, several parts of the code, such as the tree-build, have costs that scale more like particle number than the number of gravitational interactions and so do not benefit from a work-load balance based on the latter. Third, in a situation where a halo lies near a domain boundary, the optimal work-load balance will often divide the densest, shortest timestep, particles into separate processor domains. This will induce extra communication between the domains as these particles interact with each other. Thus, although this better subdivides the work, in our specific simulation it creates more overhead than it is worth.

\subsubsection{Shared Memory Locking}
\label{sec:locking}

The shared memory parallelization of the code has been improved to the point where \mpgadget~performs optimally with $>28 $ OpenMP ranks per MPI rank (in practice, we use $1$ MPI rank per processor socket).
Achieving this level of shared memory scalability required both parallelizing additional areas of the code (essentially all loops are now OpenMP parallel), and reducing the amount of code which requires taking a lock. In service of the latter goal, we have mostly replaced the per-particle spinlocks used in \cite{Feng:2016} with atomic instructions. In most cases the atomic operations supported by OpenMP's atomic update directive and reduction clauses were sufficient. However, in some places we needed to perform a non-trivial calculation before updating some particle property. In these situations we make use of the atomic\_compare\_exchange compiler built-in\footnote{Our remaining spinlock protects modification of a 128 bit structure storing the minimum particle ID in a halo during FOF table construction: we attempted an atomic implementation using the cmbxchg16b instruction but found it substantially slower than the (mostly uncontended) lock.}.


\section{Results}
\label{sec:results}

In this Section we present results from the simulation, focusing on the galaxy population. While we make comparisons to existing observations, our main goal is to make predictions for the new data which should soon become available from JWST.

\begin{figure}
\centering
  \includegraphics[width=0.45\textwidth]{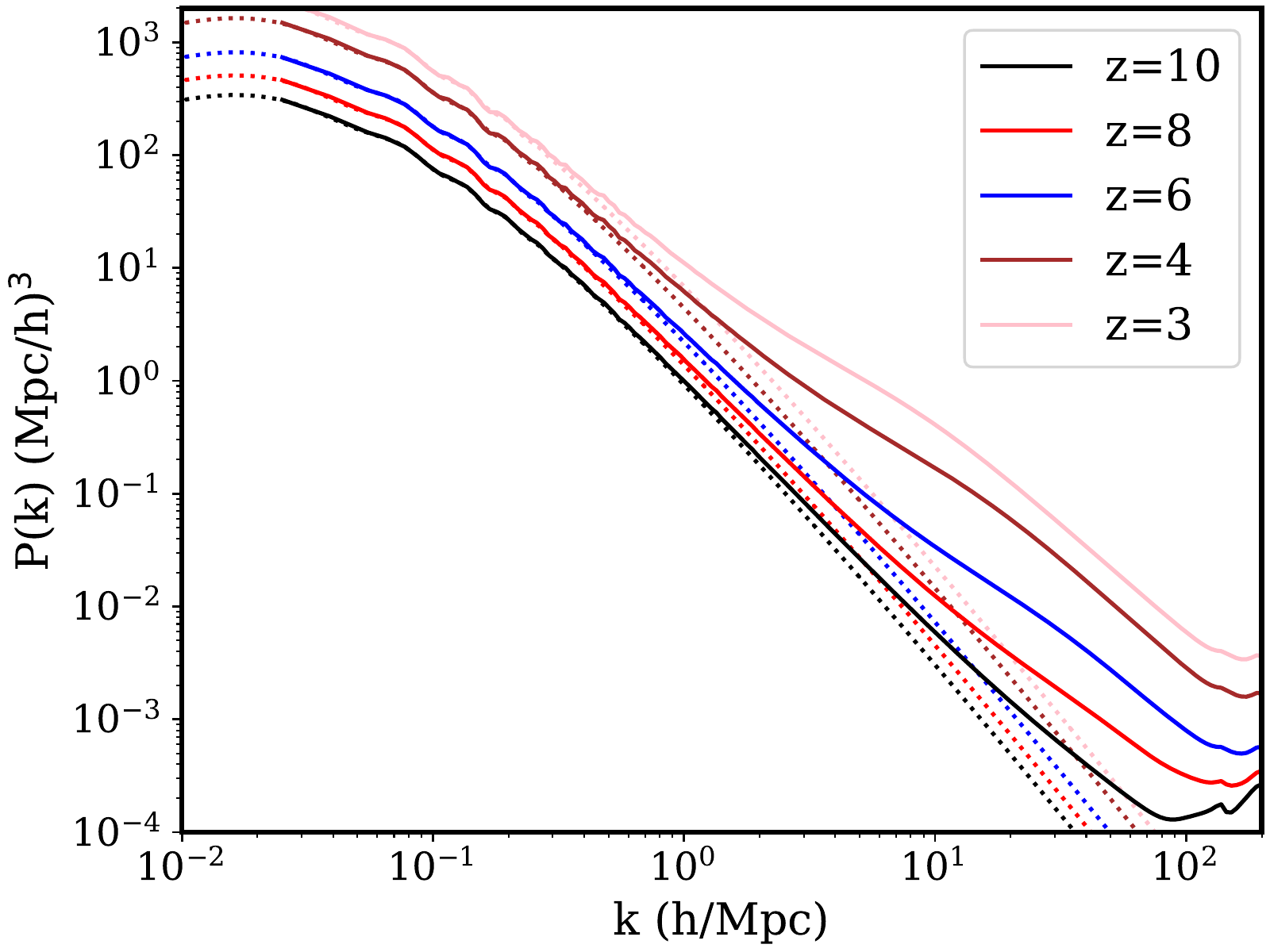}
  \caption{Matter power spectra as a function of redshift. Dotted lines show linear theory, solid lines show the simulation output.}
  \label{fig:matterpower}
\end{figure}

As a first validation of the simulation we show the matter power spectra from the simulation compared to linear theory in Figure~\ref{fig:matterpower}. The results are as expected, matching linear theory on large scales and growing more strongly on smaller scales. The scale of the mean inter-particle spacing is $k = 140 \hMpc$ and, especially at high redshift, there is a systematic visible at this scale due to the initial grid placement of cold dark matter particles.

\subsection{UV Luminosity Function}
\label{sec:uvlf}

\begin{figure*}
\centering
  \includegraphics[width=1.0\textwidth]{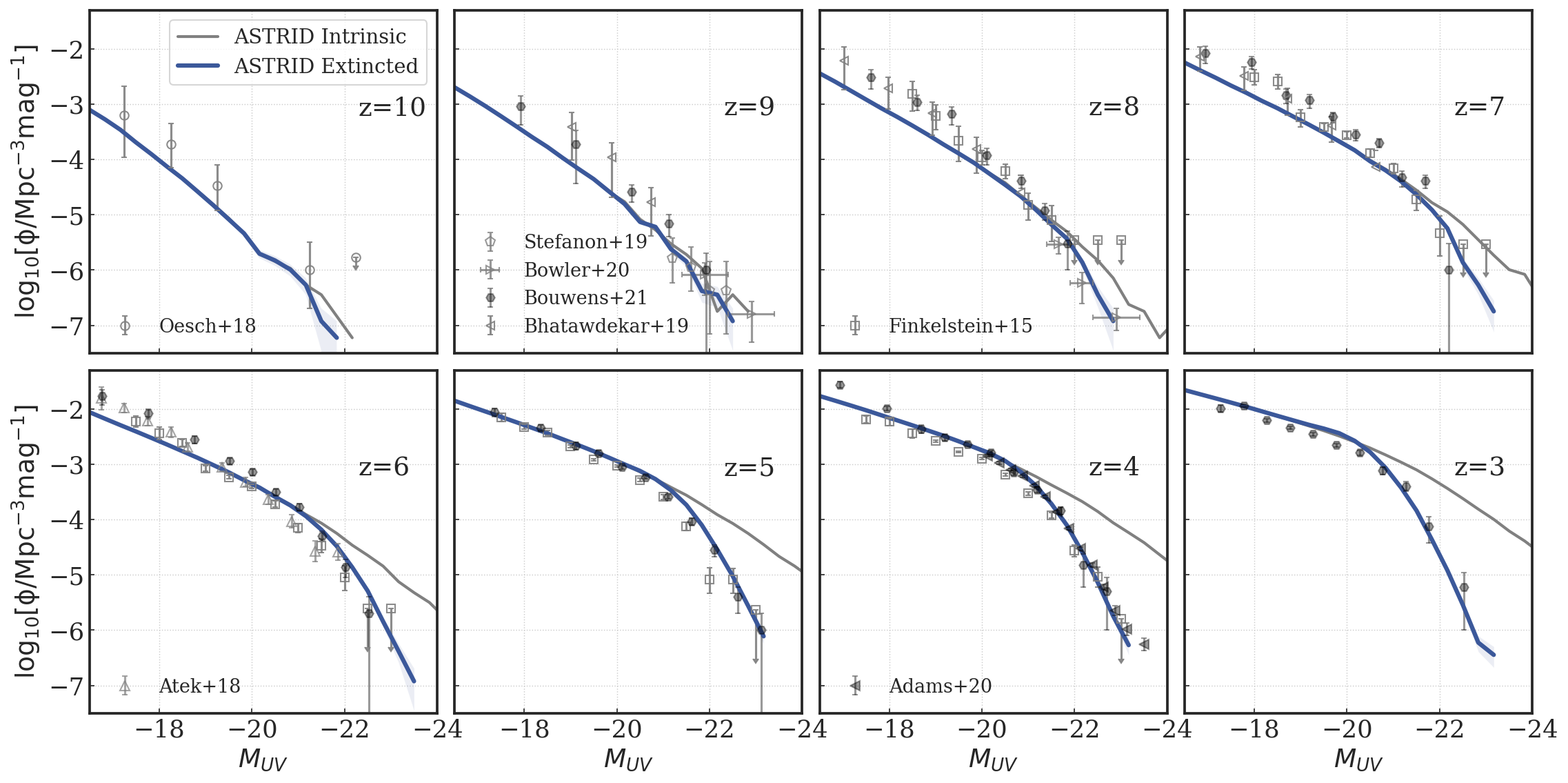}
  \caption{Galaxy UV luminosity functions. Data points with error bars give the current observational results collected from \citet{Bouwens2015,Atek2015,Atek2018,Finkelstein2015,Bhatawdekar2019,Stefanon2019,Adams2020,Bowler2020,Bouwens2021}.
  The gray dashed lines are the intrinsic UV luminosity of the FOF groups.
  The gray solid lines are the intrinsic UV luminosity for the identified \texttt{SUBFIND} galaxies.
  The blue lines are the galaxy UV luminosity corrected for dust extinction.
}
  \label{fig:UVLF}
\end{figure*}

We calculate the galactic UV luminosity function by modelling the spectral energy distribution (SED).
We assign a SED from a simple stellar population (SSP) to each star particle based on its mass, age and metallicity using the Binary Population and Spectral Population Synthesis model \citep[BPASS, version 2.2;][]{Stanway2018}, assuming a modified Salpeter initial mass function with a high-mass cut-off of $300M_\odot$.
The SED of the galaxy is taken as the sum of the SEDs of each of its star particles.

As in \citet{Wilkins2017}, we model the dust attenuation of galaxies by relating the density of metals along a line of sight to the UV-band dust optical depth $\tau_{\rm UV}$. For each star particle in the galaxy, we calculate $\tau_{\rm UV,\ast}$ as
\begin{equation}
\label{eq:dust_extinction}
 \tau_{{\rm UV,}\ast}= -\kappa \Sigma(x, y, z) \left(\frac{\lambda}{5500\text{\normalfont\AA}}\right)^{\gamma}.
\end{equation}
where $\Sigma(x, y, z) = \int_{z'=0}^{z} \rho_{\rm metal}(x,y,z')dz'$ is the metal surface density at the position of the star particle, along the z-direction line of sight, and $\kappa$ and $\gamma$ are free parameters.
Here we use $\kappa=10^{4.1}$ and $\gamma=-1.0$, which are calibrated against the observed galaxy UV luminosity function at redshift $z=4$ and applied to all redshifts.
The total dust-attenuated galaxy luminosity is the sum of the extincted luminosities of each individual star particle.

Figure~\ref{fig:UVLF} shows the UV luminosity function (UVLF) compared to a variety of observations. We also show (dashed lines) the UVLF predictions from the \texttt{SUBFIND} objects. These are very similar to the results from the FOF catalogues, as the UV luminosity at these high redshifts is dominated by the largest galaxy in the FOF group.
Physically, the UVLF is correlated with the instantaneous star formation rate; agreement with the observations suggests that our star formation rates are realistic. Comparing the results from \astrid~to observations shows that agreement is reasonable for most redshifts. Notice that the agreement in the regime where dust extinction is important is particularly good: dust extinction is calibrated at $z=4$, but the good agreement for other redshifts represents a genuine predictive success for the model, as the same dust correction is used for all redshifts. At $z < 6$, agreement with the observational data is very good even for fainter objects. For $z=7-8$, \astrid moderately under-predicts the luminosity of faint objects ($M_{UV} < -19$), although the discrepancy is at worst \citep[comparing to ][]{Bouwens2021} $2-\sigma$. At $z > 8$, we are again in agreement with observations, although the data is less constraining at these redshifts.

\begin{figure}
\centering
  \includegraphics[width=0.45\textwidth]{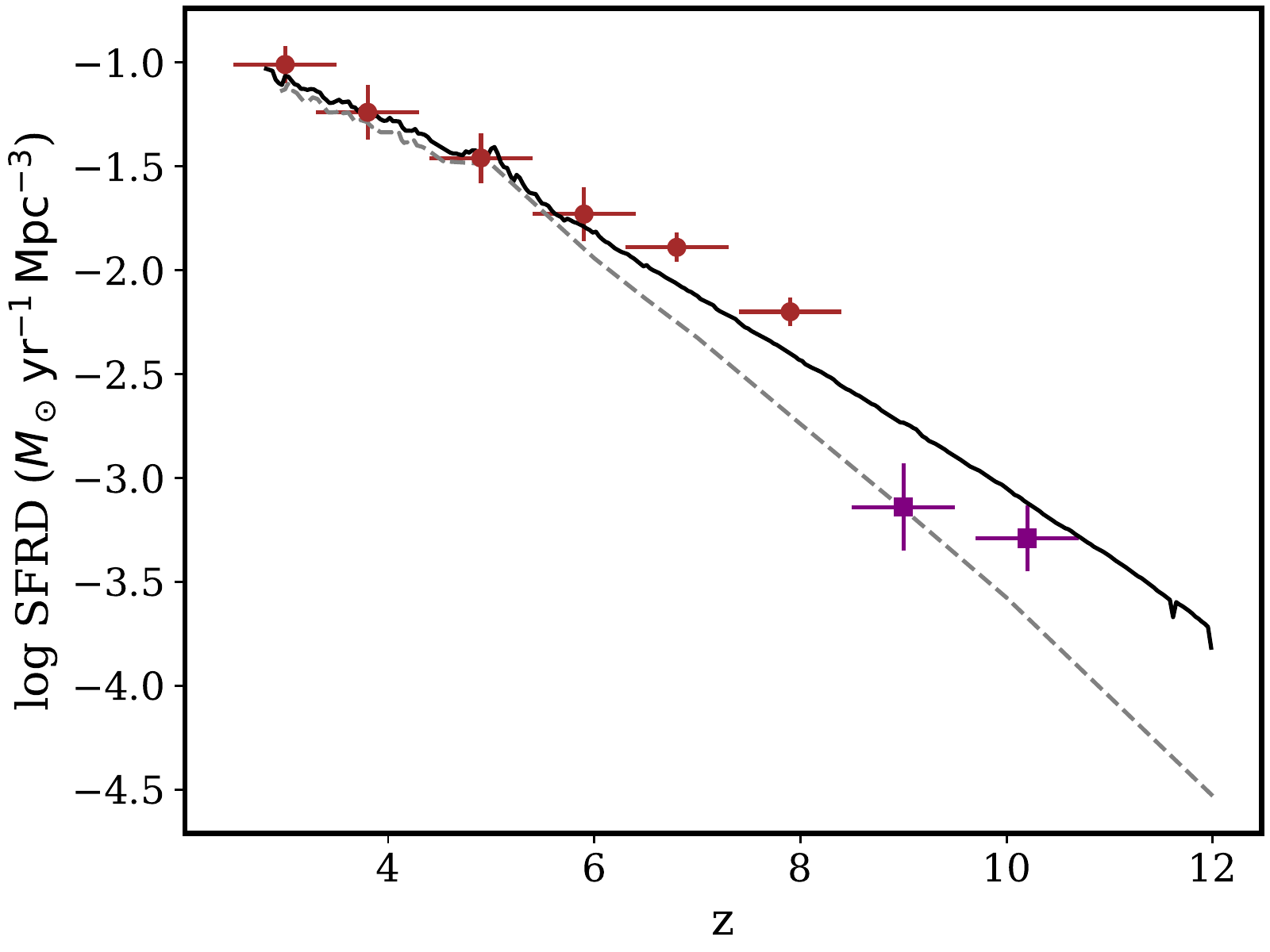}
  \caption{Star Formation Rate Density (SFRD) over time. The solid black line shows the results from the whole \astrid~simulation box, while the grey dashed line shows results for halos with a star formation rate $> 0.3 M_\odot \mathrm{yr}^{-1}$. Observational estimates are shown from \protect\cite{Bouwens:2016} (grey circles) and \protect\cite{Oesch:2014, Oesch:2018} (purple squares), with a sensitivity limit equivalent to $0.3 M_\odot \mathrm{yr}^{-1}$.}
  \label{fig:sfrd}
\end{figure}

Figure~\ref{fig:sfrd} shows a comparison with star formation rate density (SFRD) more directly. Plotted are SFRDs from the entire \astrid box, as well as star formation in halos with a star formation rate (SFR) $> 0.3 M_\odot \mathrm{yr}^{-1}$, which corresponds to current observational limits on the UV luminosity function \citep{Oesch:2018}. Current SFRD measurements at high redshift are derived by multiplying the UV luminosity by $\kappa_{UV} = 1.15\times 10^{-28} M_\odot \mathrm{yr}^{-1} \mathrm{erg}^{-1}\, \mathrm{s} \,\mathrm{Hz}$  \citep{Madau:2014}. Figure~\ref{fig:sfrd} is thus a simple transformation of Figure~\ref{fig:UVLF}. As with the UVLF, our star formation rates are in generally good agreement with observations for $z=6-3$, and at $z>9$, but low at $z=7-8$. The total star formation in the box is in better agreement with observations, but at $z > 6$ a large fraction occurs in halos below the observational luminosity limit. Figure~\ref{fig:UVLF} reveals that the discrepancy is driven by fainter objects, suggesting a higher degree of clustering in star formation or a modification of the supernova feedback model at high redshift.

\subsection{Galaxy Stellar Mass Function}
\label{sec:gsmf}

The galaxy stellar mass function, a histogram of stellar masses per unit volume, is one of the most fundamental quantities of galaxy formation theory. However, observations do not measure it directly, but deduce it from a combination of light curves and stellar population modelling.

\begin{figure*}
\centering
  \includegraphics[width=1.0\textwidth]{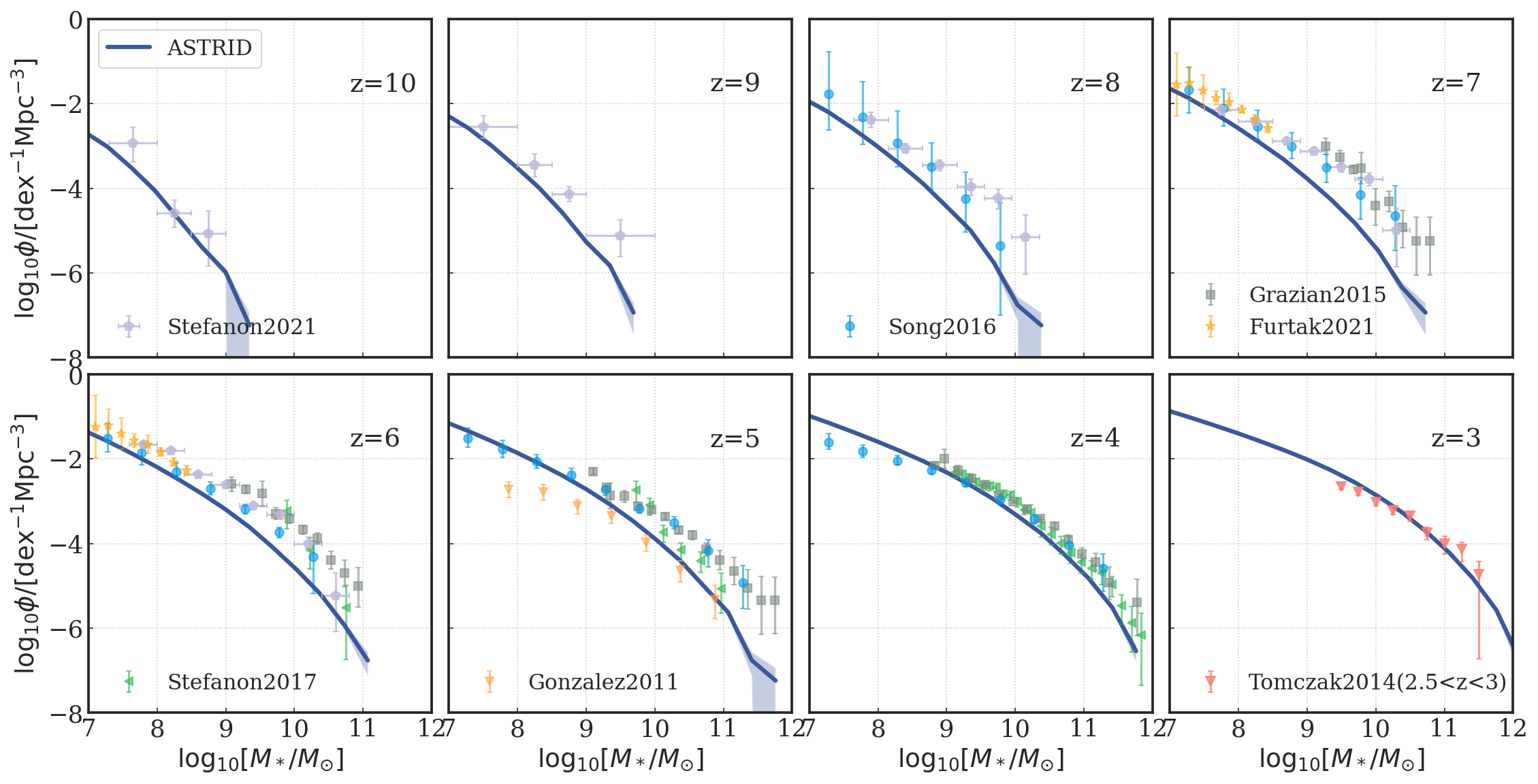}
  \caption{Galaxy stellar mass functions from \astrid~for $z=10-3$. Solid lines show the FOF halo masses, dashed lines show the mass function of galaxies in subgroups identified by running SUBFIND. Observational data collected from \citet{Gonzalez2011},\citet{Tomczak2014}, \citet{Grazian2015}, \citet{Song2016}, \citet{Stefanon2017}, \citet{Stefanon2021} and \citet{Furtak2021}. We shifted \citet{Furtak2021} results (evaluated at $z=7\sim6$) up and down by 0.15 dex for $z=7$ and $z=6$ comparison in order to account for the redshift-evolution.}
  \label{fig:GSMF}
\end{figure*}

Figure~\ref{fig:GSMF} shows our results at $z = 10 - 3$, compared to a compendium of observational data. We show the GSMF computed using stars within FOF halos, and the GSMF computed from all subgroups found by \texttt{SUBFIND}. The two GSMFs are extremely similar for $z > 4$. At these redshifts galaxies are dominated by a single large bound stellar component, and so the \texttt{SUBFIND} mass is similar to the FOF mass. A small reduction in the number of massive halos (with $M_* > 10^{11} \msun$) is visible at $z=3$, as \texttt{SUBFIND} splits the most massive objects into subcomponents. The differences, however, are smaller than the uncertainty in the observations.

We also compare to a variety of observational measurements of the GSMF. $z=4$ represents a best case: here all datasets are in good agreement. \astrid mildly under-predicts the number densities of galaxies with $M_* > 10^{10} M_\odot$, although it is within the $1-\sigma$ error of the \cite{Song2016} and \cite{Stefanon2017} surveys. It moderately over-produces galaxies with $M_* < 10^9 M_\odot$, but these faint objects are only detected in one survey, which may be incomplete.

At $z = 7-5$, \astrid appears to produce fewer galaxies with $M_* > 10^{9.5} \msun$ than observed. There is considerable scatter between different observations, reflecting the difficulty of the measurement at these high redshifts. For example, at $z=6$, the \astrid~GSMF is in good agreement with the results of \cite{Stefanon2017}, \cite{Gonzalez2011} and \cite{Song2016}, but in tension with \cite{Grazian2015} and \cite{Duncan2014}. At $z=7$, the \astrid~GSMF is lower for $M_* > 10^9 \msun$ than the midpoints of all measurements, although given the size of the error bars (and strong correlation between mass bins) the discrepancy is not particularly significant.

Stellar mass functions at $z > 4$ are deduced by combining the rest-frame UV luminosity with a luminosity per stellar mass fit on data obtained at lower redshifts, and are sensitive to the details of the spectral energy distribution (SED) assumed \citep{Furtak2021}. Note that \astrid~compares well to the observed UVLF, as shown in Figure~\ref{fig:UVLF}. While the UVLF is closer to a measurement of the star formation rate than the total stellar mass, in order for the GSMF to be low at $z=7$, star formation must also be under-predicted at some higher redshift. The UVLF is underpredicted at $z=7-8$, but substantially less than the GSMF (around $1$ survey $\sigma$, instead of $\sim 1.5$). Since the GSMF is a calibrated quantity, it is likely that the discrepancy between the simulation and the observed GSMF is influenced by inaccuracy in the calibration. With the upcoming launch of JWST it will become possible to observe these faint galaxies in the rest-frame optical and near-IR band and GSMF measurements will become more reliable.

Should the discrepancy persist observationally, possible explanations from the simulation likely come from the stellar feedback which controls the GSMF at these redshifts and masses. The largest discrepancy, occurs at $z=7$ in halos with stellar mass $10^{9.5} < M_* < 10^{10} \msun$, stellar masses and redshifts where AGN feedback has little effect. Using a test simulation with a $25$ Mpc/h box, and $550^3$ particles (thus the same mass resolution as \astrid), we found that immediately recoupling stellar wind particles to the gas (ie, allowing them to experience hydrodynamic forces), produced an increase in the GSMF which brought it into better agreement with $z=7$ observations. However, by $z=2.5$ this model  substantially over-produced stars compared to observations.




\subsection{Star Formation Rates}
\label{sec:sfr}


\begin{figure}
\centering
  \includegraphics[width=0.45\textwidth]{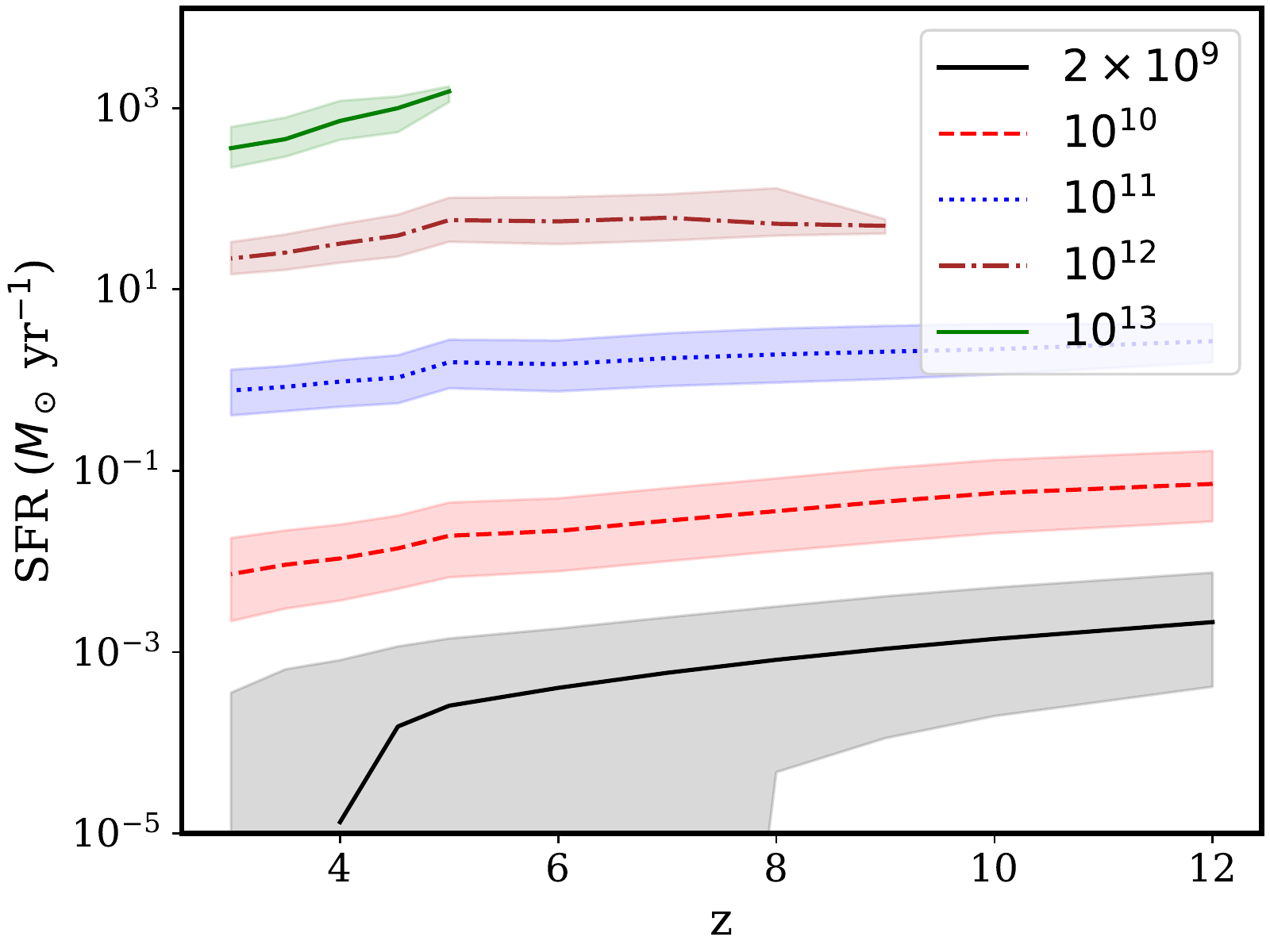}
  \caption{Star formation rates averaged over all halos in halo mass bins centered on $2\times 10^9$, $10^{10}$, $10^{11}$, $10^{12}$ and $10^{13} \msun$. Bins contain halos with $\pm 30\%$ of the central halo mass. Solid lines show median, bands show $16$ and $84$ th percentiles. The coloured bands are thus not error bars, but measures of the stochasticity of star formation. The smallest mass bin is chosen to contain on average $250$ DM particles.}
  \label{fig:avgsfr}
\end{figure}

\begin{figure}
\centering
    \includegraphics[width=0.45\textwidth]{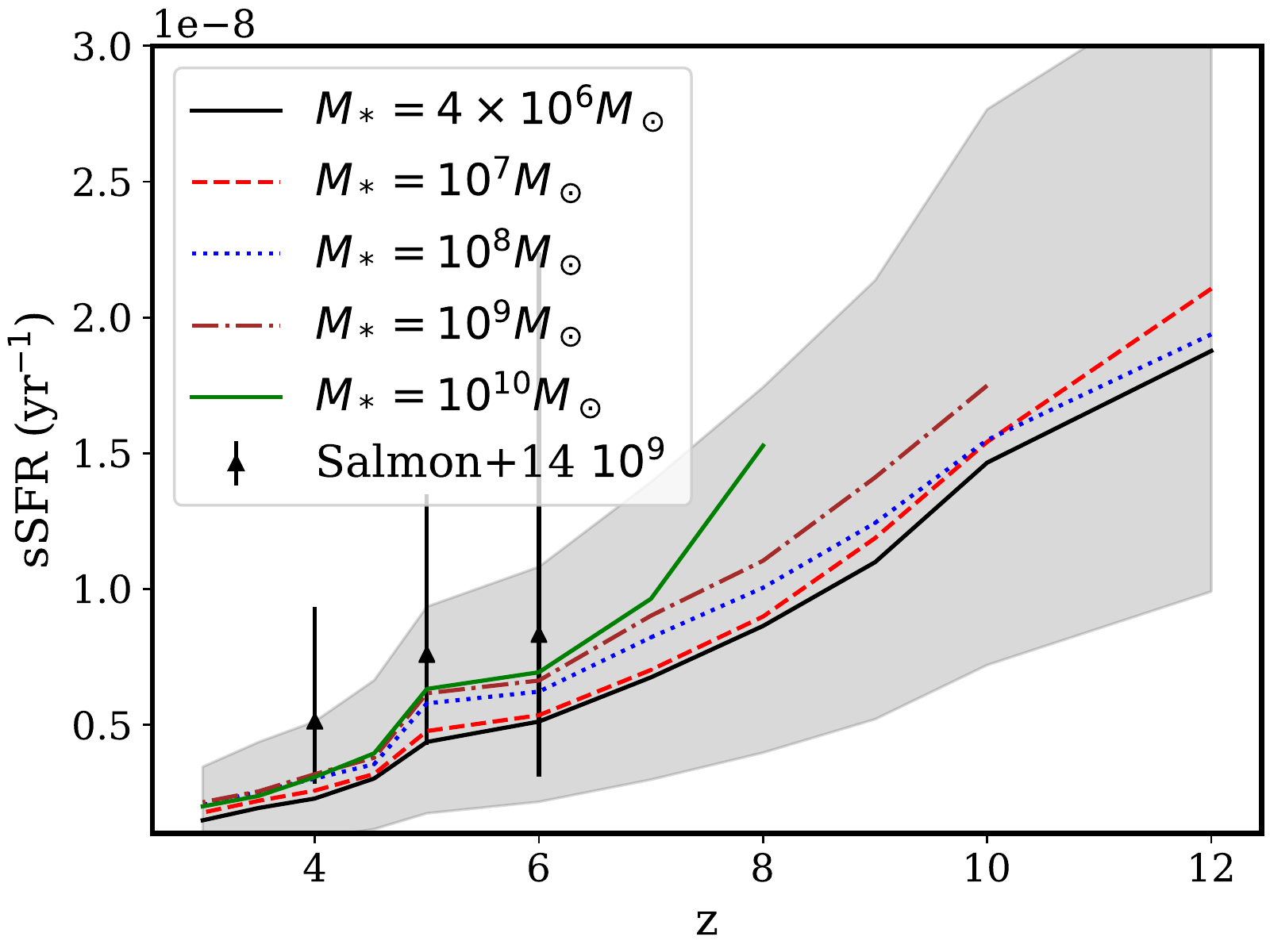}
  \caption{Specific star formation rate (star formation rate per unit stellar mass). Shown are the median of all halos in stellar mass bins centered on $M_* = 4\times 10^{6}$, $10^{7}$, $10^{8}$, $10^9$ and $10^{10} \msun$ and spanning $\pm 30$ percent. Grey band shows the $16$ and $84$ th percentile in the $M_* = 4\times 10^6 \msun$ bin. We compare to the results of \protect\cite{2015ApJ...799..183S}, based on UV and optical data.}
  \label{fig:ssfr}
\end{figure}

Figure~\ref{fig:avgsfr} shows the average star formation rate at fixed halo mass as a function of redshift. Although this quantity is not directly observable as halo masses are not accessible for this era, we nevertheless show it for physical insight. Shown is the median star formation rate in each bin, as well as the $16$th and $84$th percentiles as coloured bands. The bands thus show not statistical uncertainty, but a measure of the scatter between star formation rates in different halos. This scatter is larger as a fraction of the SFR in smaller halos. This reflects both that star formation in small halos is inherently bursty due to supernova feedback and that the number of star forming particles in small halos is lower. Average star formation rates are as expected, larger in more massive halos, and reduce moderately with decreasing redshift. This reduction is especially noticeable for low mass halos after HI reionization ($z \sim 6$), and after the start of HeII reionization ($z=4.4$). In Section~\ref{sec:sfrreion} we will examine the effect of reionization on star formation rates more directly.

We do not show a comparison to observational estimates of star formation rates \citep[see][for a review]{2013ApJ...770...57B}. The underlying observations are generally UV luminosity functions, combined with dust correction estimates. We directly predict UV luminosity functions from the simulation in Figure~\ref{fig:UVLF} and so little information would be added by comparing our simulation to star formation rates derived from this same observational data. Directly predicting the observational data from the simulation is more reliable as the metallicity of the galaxy is observationally uncertain but available in the simulation, allowing better estimates of dust extinction.

Finally, note that halo masses are not observed quantities but estimated using abundance matching and thus observational results which compare to them implicitly rely on dark matter only simulations. \astrid~has the same box size as the MultiDark simulation \citep{2016MNRAS.457.4340K} used by \cite{2019MNRAS.488.3143B} for abundance matching star formation histories, but includes substantially higher fidelity models of the physical processes of galaxy formation.

Figure~\ref{fig:ssfr} shows the specific star formation rate (the star formation rate per unit stellar mass, sSFR) for a variety of stellar mass bins. Our results show that star formation rate per unit stellar mass is approximately constant as a function of stellar mass, but increasing with redshift. Between $z=9$ and $z=5$ the median sSFR in galaxies with $M_* \leq 10^7 \msun$ is about $20\%$ lower than for galaxies with $M_* \geq 10^8 \msun$, which may be due to a suppression of star formation in small halos during reionization. We examine this further in more detail in Section~\ref{sec:sfrreion}.

The overall redshift trend is for a decreasing specific star formation rate, from $1.5 \times 10^{-8}$ yr$^{-1}$ at $z=10$ to $3 \times 10^{-9}$ yr$^{-1}$ at $z=4$. This is consistent with the expectation of the halo model: the concentration and thus internal density of a halo is proportional to the background density at the time it virializes. Halos of a given mass at lower redshift have virialized later and thus have a lower internal density. Our predictions are in reasonable agreement with the observational star formation measurements of \cite{2015ApJ...799..183S}, although observational errors at high redshifts remain relatively large. We also agree with the abundance matching model of \cite{2013ApJ...770...57B} (not shown), which is influenced by lower redshift data through N-body simulations.

\subsection{Mass Fractions}

\begin{figure}
\centering
  \includegraphics[width=0.45\textwidth]{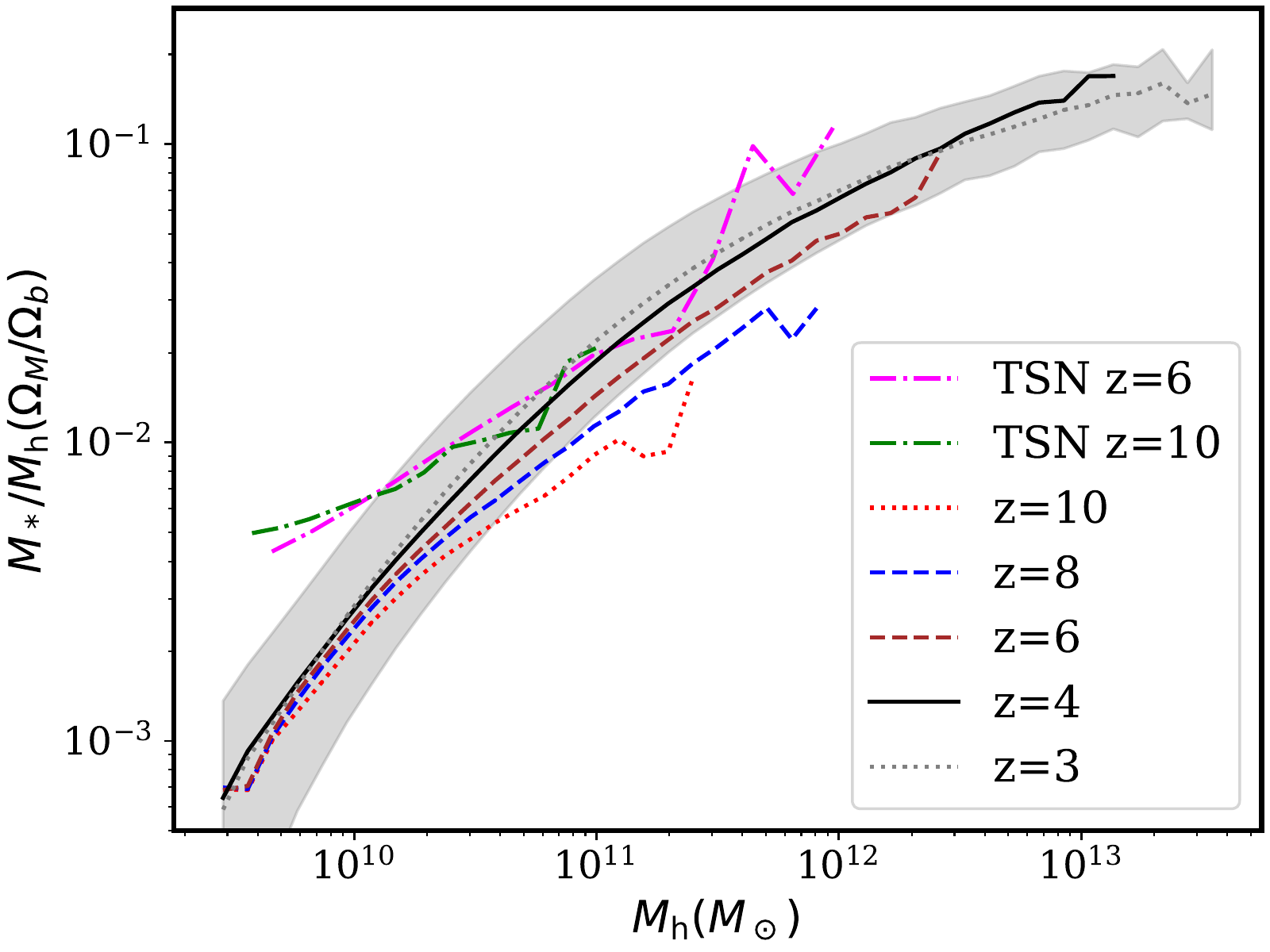}
  \caption{Median stellar mass divided by halo mass as a function of halo mass, for $z=10-3$. Each halo mass bin must contains at least $10$ halos, to reduce statistical noise. Lines marked TSN are from the THESAN simulation \protect\citep{2021arXiv211000584K}. The grey band shows the $16$th and $84$th percentiles in the lowest redshift snapshot at $z=3$.}
  \label{fig:smhm}
\end{figure}

\begin{figure*}
\centering
  \includegraphics[width=0.45\textwidth]{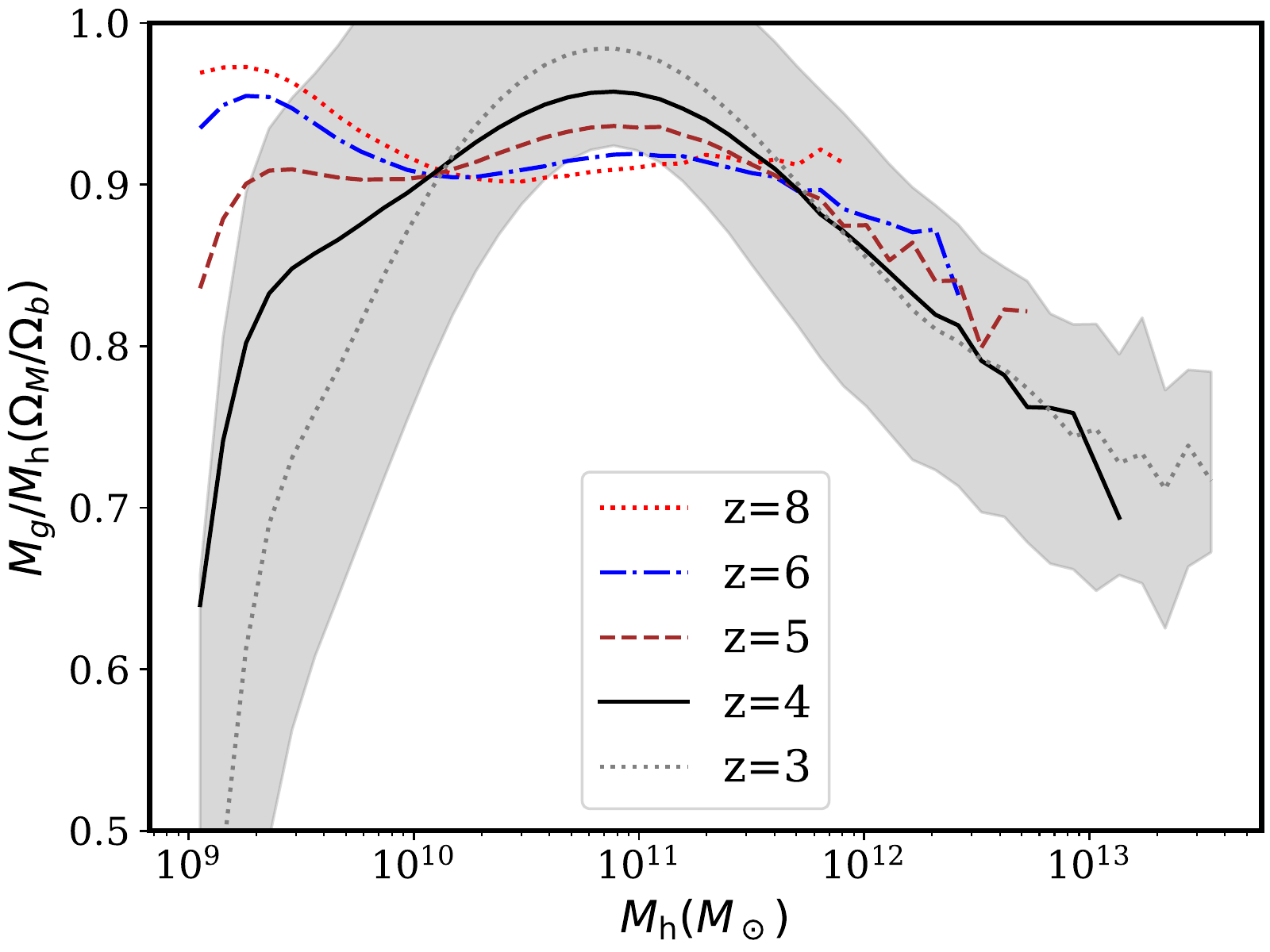}
  \includegraphics[width=0.45\textwidth]{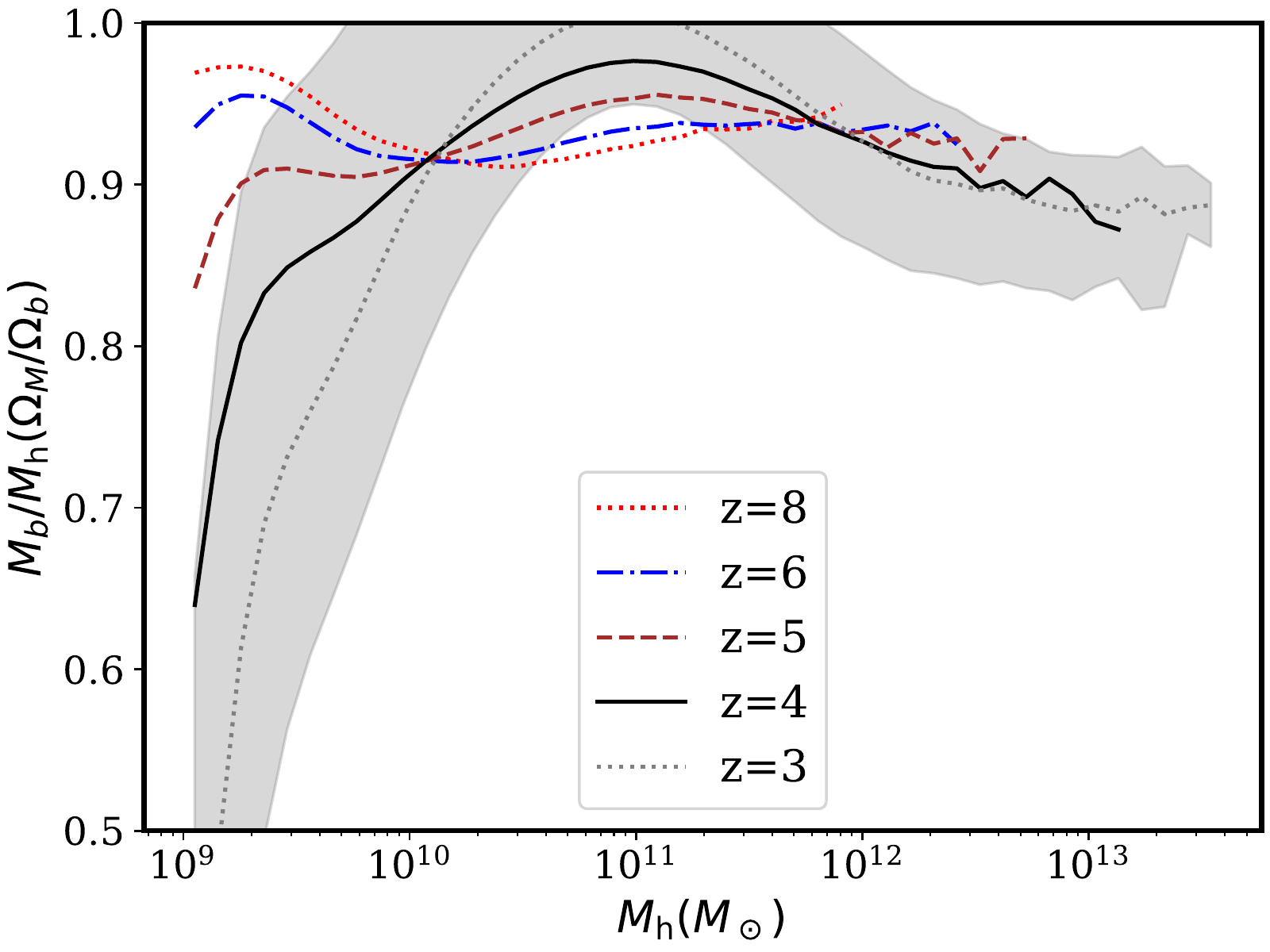}
  \caption{Median gas mass divided by halo mass as a function of halo mass, at $z=10-3$. Each halo mass bin must contains at least $10$ halos, to reduce statistical noise. Grey band shows the $16$th and $84$th percentiles in the lowest redshift snapshot at $z=3$. (Left) Gas fraction only. (Right) Baryon (gas, star and black hole) fraction. }
  \label{fig:gmhm}
\end{figure*}

\begin{figure}
\centering
  \includegraphics[width=0.45\textwidth]{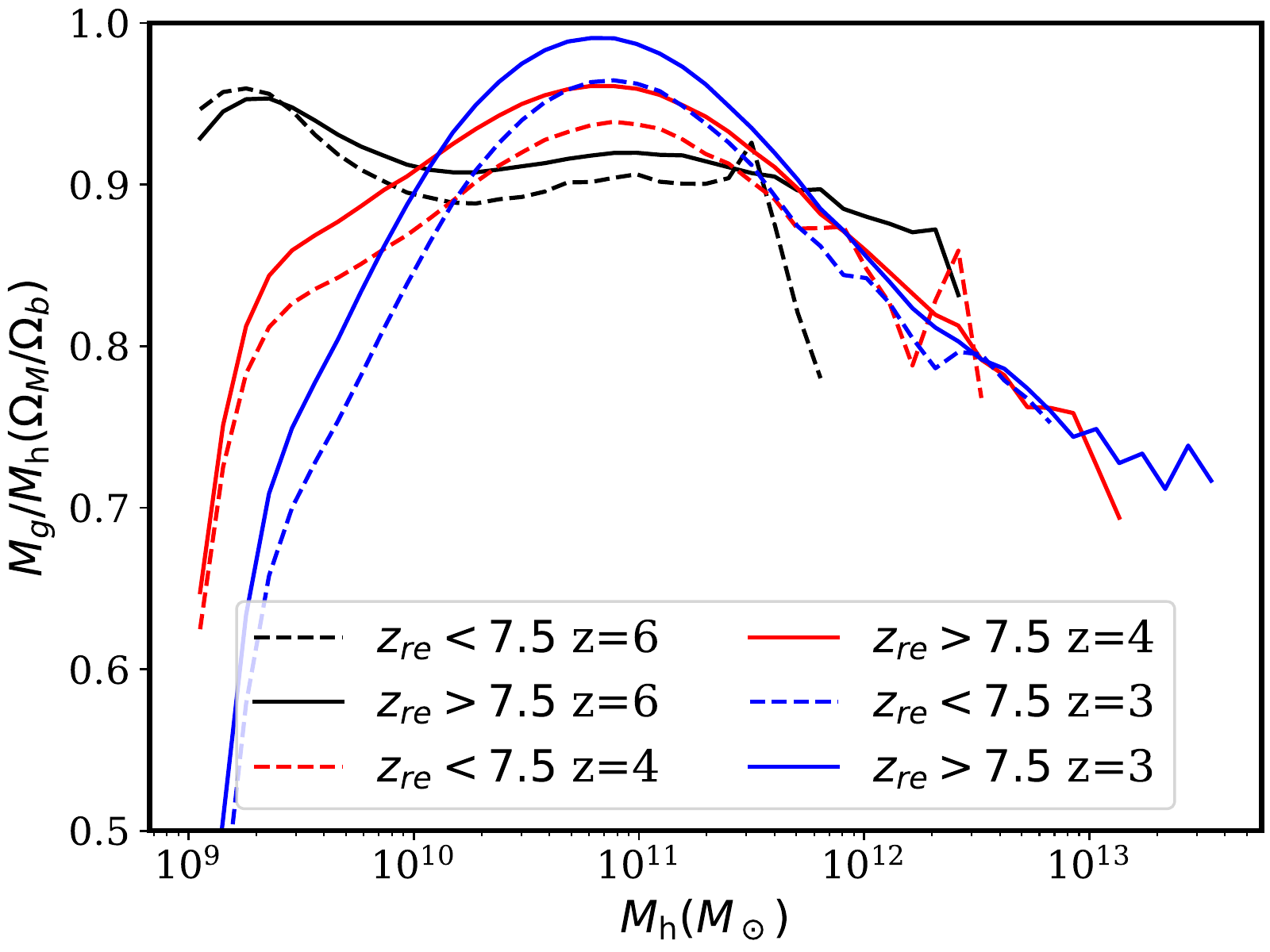}
  \caption{Median gas mass divided by halo mass as a function of halo mass. Each halo mass bin must contains at least $10$ halos, to reduce statistical noise. Shown is the gas fraction in halos that reionized before and after $z=7.5$, the midpoint of reionization in our model. Halos which reionized later have a lower gas fraction.}
  \label{fig:gmhm_reion}
\end{figure}

Figure~\ref{fig:smhm} shows the stellar mass-halo mass relation, as well as an estimate of the scatter at the $z=3$. The scatter at higher redshift is similar. For low halo masses the stellar mass - halo mass relation is redshift invariant. For $M > 10^{10} \msun$, halos with similar masses at higher redshifts have lower stellar mass, reflecting that they are only recently assembled and have not yet had an opportunity to form stars.

Our results are in good agreement with models derived from abundance matching \citep{2019MNRAS.488.3143B}. Figure 6 of \cite{2014MNRAS.445..175G} shows that our results agree with the stellar mass fraction predicted by the Illustris simulation at $z = 3-4$, demonstrating that our feedback model is behaving as expected. Note that at $z=0$, the stellar mass fraction shows a peak around halos of mass $10^{12} \msun$ \cite[e.g.][Figure 11]{2018MNRAS.475..648P}. The peak is driven by AGN-driven quenching star formation in high mass halos. At $z=3$ galaxies have generally not yet quenched, and so our results do not show a similar peak.

We compare also to the results from the recent \texttt{THESAN} radiative transfer simulations \citep{2021arXiv211000584K}. The comparison is limited to $z>6$ as \texttt{THESAN} has so far only reached $z=5.5$. Our results are similar within the scatter of the stellar-mass/halo mass relation. \texttt{THESAN} produces more stars in its most massive halos, but their relatively small box ($65$ Mpc/h as compared to our $250$ Mpc/h) suggests this is likely cosmic variance. There is a significantly larger fraction of stars in small halos with $M_h < 10^{10} M_\odot$ in \texttt{THESAN}. Furthermore, while the stellar-mass/halo-mass relation evolves modestly with redshift in \astrid, for a fixed halo mass it is constant in \texttt{THESAN}. Since both these features occur for $z > 10$, it is unlikely they are due to differences in the modelling of reionization between the two simulations. Our feedback models are also very similar at these redshifts. There are two plausible modelling differences between the simulations which could explain these effects. The first is our inclusion of a star formation correction for molecular hydrogen, which suppresses star formation in metal poor gas (such as is found at high redshift). However, Figure~\ref{fig:stellar_metal} shows that the metallicity at $M_h = 10^{10} M_\odot$ does not evolve significantly with redshift. Another possibility is that these changes are due to differences in our cosmological model, which includes the relative velocities between baryons and dark matter, as well as the Hubble drag from radiation. We will investigate these effects further in future work.

Figure~\ref{fig:gmhm} shows a gas mass-halo mass relation, in units of the cosmological baryon fraction, as well as an estimate of the scatter at $z=3$ (the scatter at higher redshift is again similar). For $z \geq 6$, the gas fraction in halos is approximately redshift independent, and close to, but lower than, the cosmological average. We have also computed the fraction of the total mass in baryons: gas, stars and black holes, shown in the right panel of Figure~\ref{fig:gmhm}. The baryon fraction is flatter than the gas fraction for the heavily star-forming halos with $M> 10^{12}\msun$, and at $z < 5$ the baryon fraction approaches the cosmological average. However, at $z \geq 6$ the baryon fraction is still relatively low. We speculate that this may be an effect of the separate transfer functions we implemented for baryons and cold dark matter. At $z \sim 10$ the baryon power spectrum is $10\%$ lower than the dark matter power spectrum, mirroring the decrease in the gas fraction.

Around $z=5$, the gas fraction drops for $M_{\rm h} < 10^{10} \msun$ and rises for $M_{\rm h} > 10^{11} \msun$. Helium reionization begins at $z=4.4$, making it tempting to ascribe the low gas fraction for $M_{\rm h} < 10^{10} \msun$ at $z \leq 4$ to the extra heating from the reionization process. However, we have checked that the gas fractions in helium reionized halos and non-reionized halos are extremely similar. Interestingly, the gas fraction is affected by hydrogen reionization as shown in Figure~\ref{fig:gmhm_reion}. Regions of the box which reionize before $z = 7.5$ have larger gas fractions than regions which reionize after $z=7.5$, with the effect larger at lower redshift. A plausible explanation is that heating during reionization creates a pressure shock which temporarily reduces gas accretion, before dissipating. Regions which reionize earlier have had more time for the pressure shock to dissipate and thus are able to accrete at a higher rate. Note that a model in which non-reionized regions have larger accretion simply because of their lower gas temperatures would produce a trend opposite to that seen, with early reionization leading to lower gas fractions.

Another plausible explanation is that, in our reionization model, $z_{re}$ is correlated with overdensity on Mpc-scales. It is possible that this higher overdensity causes intrinsically higher gas fractions, through higher accretion from nearby filaments. Note however that the overdensity is computed over smoothed $1$ Mpc/h regions, which is much larger than the virial radius of the halos.
If the initial overdensity is the cause of the effect, we would expect a dependence of gas fraction on halo concentration: halos which formed earlier, in a higher overdensity, would reionize earlier and have a higher gas fraction. While we did not perform an NFW fit, we defined an effective halo radius using the diagonal elements of the moment of inertia $I_{i j}$ of the FOF halo as:
\begin{align}
    R_{eff} &= \left[(I_{0 0} + I_{1 1} + I_{2 2})/(3 M_h) \right]^{1/2} \\
    I_{i j} &= \Sigma m r_i r_j \\
    M_h &= \Sigma m
\end{align}
where the sum is over all particles in the halo and $r_i$ is position relative to the halo center of mass. For fixed halo mass, the gas fraction had no correlation with $R_{eff}$, and the dependence on $z_{re}$ persisted, indicating that halo internal density is not driving this effect.


\subsection{Stellar and Gas Metallicity}

\begin{figure}
\centering
  \includegraphics[width=0.45\textwidth]{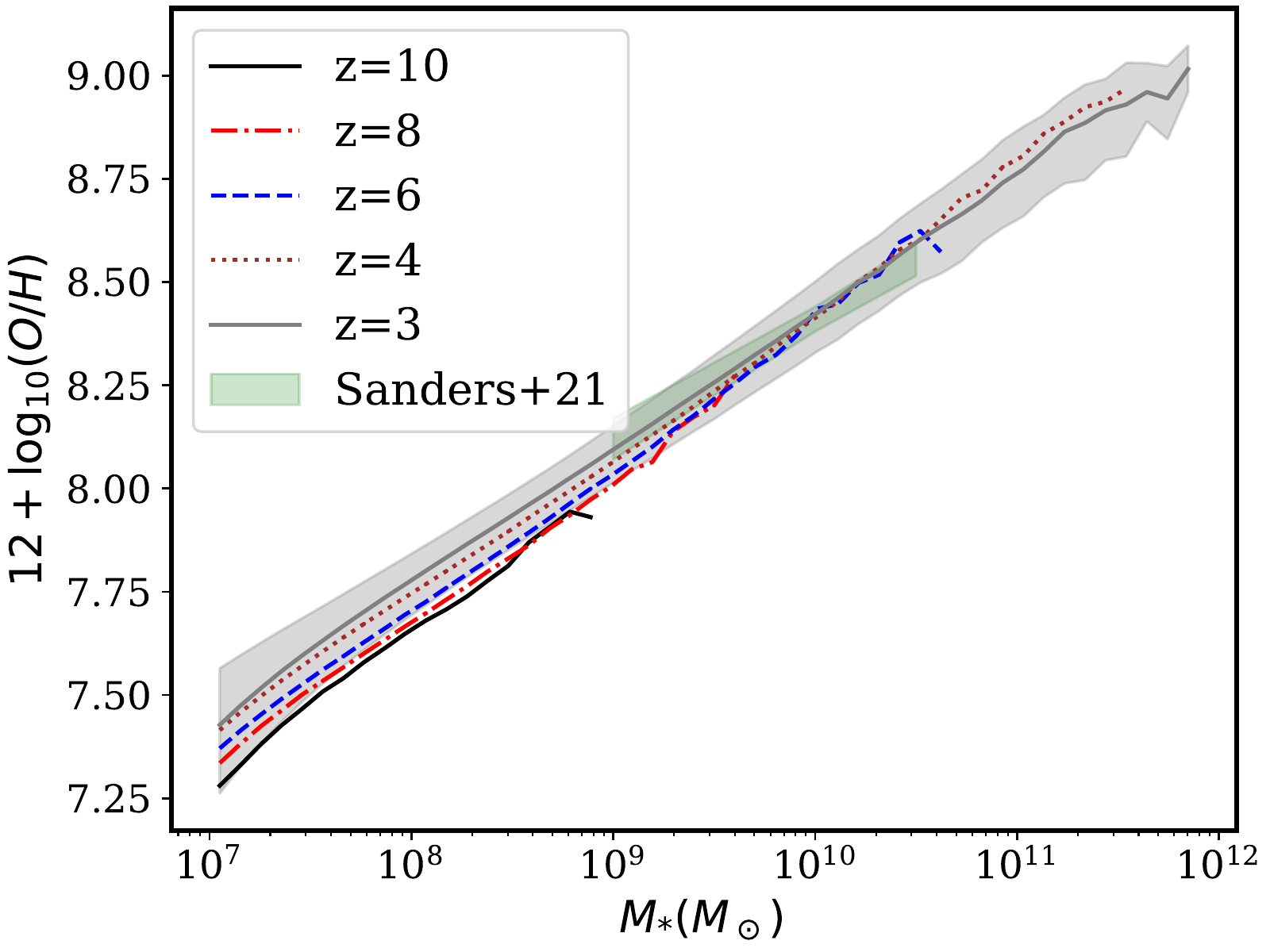}
  \caption{Stellar oxygen abundance as a function of stellar mass. Shown is the oxygen number abundance per unit hydrogen abundance in stars. The hydrogen mass abundance is assumed to be $0.76\times M_*$. Data is a power law fit to a stack of galaxies centered on $z=3.3$ \protect\citep{2021ApJ...914...19S}.}
  \label{fig:stellar_metal}
\end{figure}


Figure~\ref{fig:stellar_metal} shows the stellar O/H abundance as a function of stellar mass. Note that the normalisation of the mass metallicity relation is observationally uncertain \citep{2019MNRAS.484.5587T} and so the most reliable comparison is to the slope. Nevertheless, our simulation is in good agreement with the observations of \cite{2021ApJ...914...19S} in both normalisation and slope, in the limited mass range where they currently overlap. The mass-metallicity relation in our simulation forms a single unbroken power law over $5$ decades of galaxy stellar mass. Redshift evolution is small, but not zero, with the normalisation of the mass-metallicity relation increasing over time as extra metal mass is formed.

\subsection{Effect of Hydrogen Reionization on Star Formation Rates and Gas Fractions}
\label{sec:sfrreion}

\begin{figure}
\centering
  \includegraphics[width=0.45\textwidth]{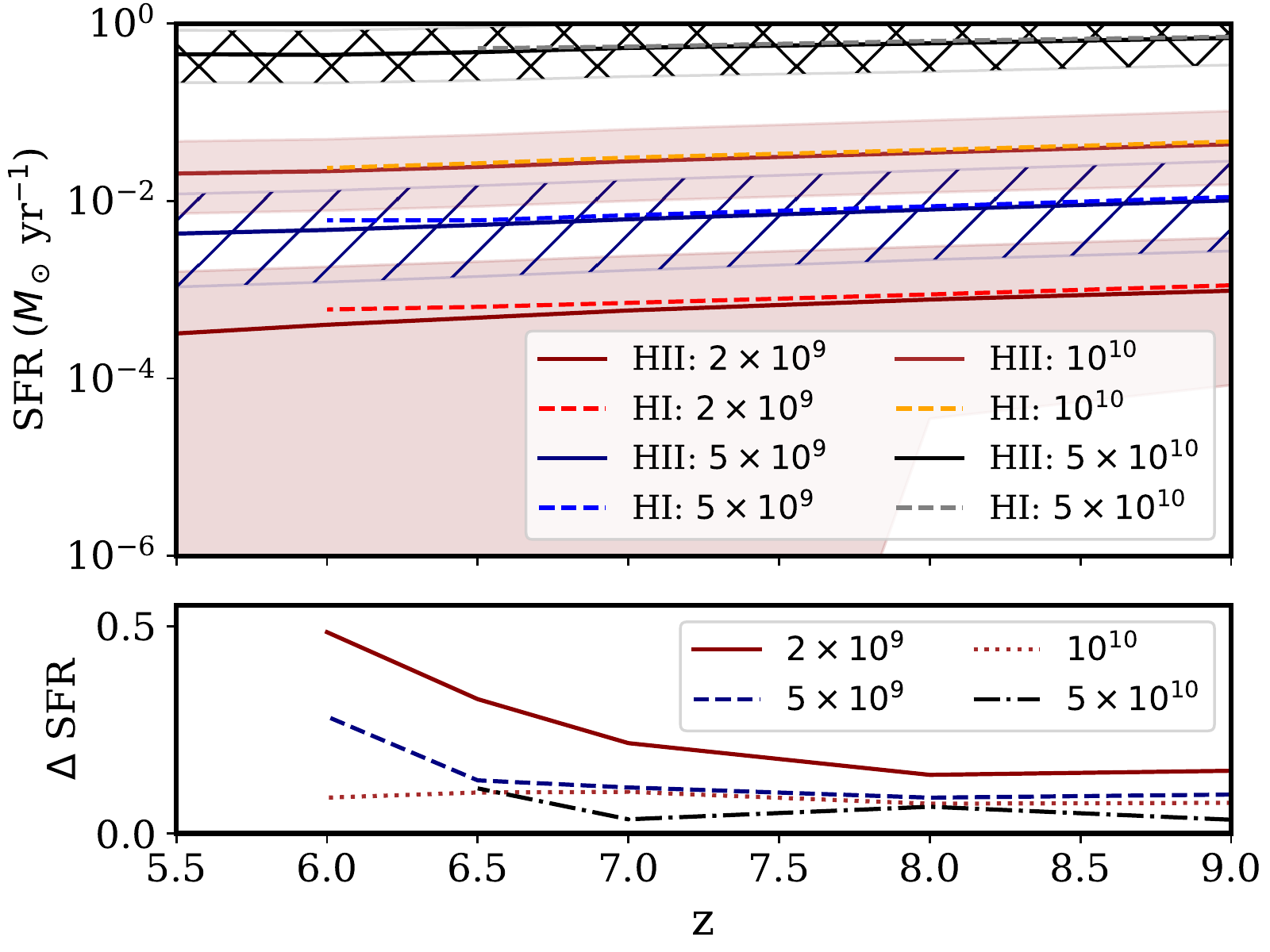}
  \caption{Star formation rates for reionized (solid) and non-reionized (dashed) halos. Bands and hatched areas show $16$ and $84$ th percentiles. We average over all halos in halo mass bins centered on $10^{10}$, $10^{11}$, $10^{12}$ and $10^{13} M_\odot$. Lower panel shows the ratio between the star formation rates in ionized and non-ionized halos. }
  \label{fig:avgsfr_reion}
\end{figure}

Figure~\ref{fig:avgsfr_reion} shows the effect of HI reionization on the star formation rate (SFR) in each halo. Note that we do not include an instantaneous heating effect from the passage of the ionization fronts \citep{2019ApJ...874..154D}, so this probably underestimates the magnitude of the effect. Nevertheless, we find that low mass halos experience a marked reduction in star formation rate ($50\%$ in the lowest mass bin) after reionization, especially for $z<7$ \citep[as also found by][]{2018MNRAS.480.1740D}. The redshift dependence may be due to the increasing UVB amplitude. We also show the scatter: note this is not the sampling uncertainty, but the width in the distribution of SFRs. Although the scatter is large, representing the bursty nature of low-mass star formation, the change in the star formation rates is stable with redshift, suggesting that the effect of reionization on star formation in our simulation is persistent. There is no effect for halos with $M_{\rm h} > 10^{10} \msun$, a median stellar mass of $M_* = 2 \times 10^7 \msun$. Measurements of the GSMF already exist for these stellar masses (see Figure~\ref{fig:GSMF}), although at $z > 6$ the uncertainties are currently larger than the expected impact of reionization. Nevertheless, there is the possibility that it may be possible to measure this effect with JWST.

We have also checked whether the redshift of inhomogeneous helium reionization in the simulation altered star formation rates and found minimal change. Although helium reionization injects substantial thermal energy, it does so only once for each gas particle. In our simulation, star formation is regulated by feedback, and the cumulative injection of feedback energy over time dominates the injection from helium reionization.\footnote{Our feedback model injects energy kinetically, but ultimately this produces thermal shocks.}

\section{Conclusions}

In this paper, we have introduced the \astrid~simulation. \astrid is a large-scale, $250 \mpch$ box, simulation with $2\times 5500^3$ particles. It thus contains a statistical sample of halos equivalent to  MultiDark \citep{2016MNRAS.457.4340K}, yet is able to resolve galaxy formation at a level comparable to Illustris-TNG 100 \citep{2018MNRAS.475..648P}. It has been run from $z=99$ to $z=3$\footnote{During the preparation of this paper we continued to run the simulation and it has now reached $z=2.5$.}. A particular focus has been on modelling the high redshift Universe and \astrid~includes models for inhomogeneous hydrogen and helium reionization, baryon relative velocities and massive neutrinos. We have briefly summarised the implemented models, and the technical choices we took when developing the simulation code.

We have shown that \astrid~reproduces current observations of galaxies at $z=10-3$, including the UV luminosity function and the galaxy stellar mass function. We also computed stellar and gas mass fractions, metallicities and star formation rates. We showed that stellar metallicities and specific star formation rates were in good agreement with the limited number of observations currently available. Stellar mass fractions were similar to other simulations, but lower than \texttt{THESAN} for halo masses $< 10^{10} M_\odot$. We showed that hydrogen reionization has a strong effect on star formation rates and gas fractions. Our companion paper \citep{Ni:2021inprep} examines the statistics of black holes.

\astrid contains an unusually large number of high redshift halos, which will enable several related investigations. In future work we will make mock galaxy observations for JWST, mock black hole merger catalogues for LISA, as well as mock X-ray catalogues and quasar spectra.

\section*{Acknowledgements}

\astrid was run on the Frontera facility at the Texas Advanced Computing Center.
We acknowledge helpful discussions with Steve Wilkins and Aswin Vijayan.
SB was supported by NSF grant AST-1817256 and would like to thank his wife, Priya Bird.
TDM and RACC acknowledge funding from
the NSF AI Institute: Physics of the Future, NSF PHY-2020295,
NASA ATP NNX17AK56G, and NASA ATP 80NSSC18K101. TDM acknowledges additional support from  NSF ACI-1614853, NSF AST-1616168, NASA ATP 19-ATP19-0084,
  and NASA ATP 80NSSC20K0519, and RACC from NSF AST-1909193.
We are grateful for the work of Astrid Lindgren, the author of Pippi Longstocking, after whom this simulation is named.

\section*{Data Availability}

The code to reproduce the simulation is available at \url{https://github.com/MP-Gadget/MP-Gadget} in the \texttt{asterix} branch, and continues to be developed. Text file forms of the data presented here as well as scripts to generate the figures are available at \url{https://github.com/sbird/asterix_intro}. Halo catalogues and snapshot particle tables are available on reasonable request to the authors. The particle tables are about $20$ TB each and the halo catalogues are $\sim 2$ TB.




\bibliographystyle{mnras}
\bibliography{example} 



\bsp	
\label{lastpage}
\end{document}